\documentclass[a4paper,twoside,10pt,english,numbers=noenddot]{scrbook}

\usepackage[english]{babel}
\makeatletter
\def\month@ngerman{\ifcase\month \or Januar\or Februar\or M\"arz\or April\or Mai\or Juni\or Juli\or August\or September\or Oktober\or November\or Dezember\fi}
\def\month@english{\ifcase\month \or January\or February\or March\or April\or May\or June\or July\or August\or September\or October\or November\or December\fi}
\makeatother
\usepackage{hyphenat} 

\usepackage[T1]{fontenc}
\usepackage{helvet} 
\usepackage{indentfirst}
\usepackage[keeplastbox]{flushend} 

\usepackage[usenames,dvipsnames]{xcolor} 
\usepackage{graphicx} 
\usepackage{wrapfig} 
\usepackage{tikz}
\usepackage{rotating}

\usepackage{ifthen}
\usepackage{forloop}
\usepackage{etoolbox}

\usepackage[fleqn]{amsmath} 
\usepackage{amssymb} 
\usepackage{amsfonts} 
\usepackage{amstext} 
\usepackage{wasysym}
\usepackage{mathrsfs} 
\usepackage{mathtools}
\usepackage{upgreek}
\usepackage{siunitx}
\usepackage{esint} 
\usepackage{cancel} 
\usepackage{empheq}
\allowdisplaybreaks

\usepackage{multicol} 
\usepackage{multirow} 
\usepackage{dcolumn}
\usepackage{tabularx} 
\newcolumntype{L}[1]{>{\raggedright\arraybackslash\hsize=#1\hsize}X}
\newcolumntype{R}[1]{>{\raggedleft\arraybackslash\hsize=#1\hsize}X}
\newcolumntype{C}[1]{>{\centering\arraybackslash\hsize=#1\hsize}X}
\usepackage{dblfloatfix} 

\usepackage{enumitem} 
\setlist{nosep} 

\usepackage[a4paper,portrait]{geometry}
\newlength{\TwoColumnWidth}
\newlength{\OneColumnWidth}
\usepackage[headsepline]{scrlayer-scrpage}

\usepackage[hang,bottom]{footmisc}
\deffootnote{\footnotemargin}{0pt}{%
	\textsuperscript{\thefootnotemark}
}
\usepackage[singlelinecheck=false,font=small,labelfont=bf,format=plain]{caption} 
\usepackage[singlelinecheck=true,font=small,labelfont=bf,format=plain]{subfig}
\setkomafont{sectioning}{\rmfamily\bfseries}
\setkomafont{descriptionlabel}{\rmfamily\bfseries}

\usepackage[subfigure]{tocloft} 
\addto\captionsenglish{

}
\makeatletter
  \renewcommand*{\@pnumwidth}{20pt} 
  \renewcommand*{\@tocrmarg}{30pt plus 5pt minus 0pt} 
  \renewcommand*{\@dotsep}{4} 
\makeatother

\newcounter{chapterappendixcounter}[chapter]

\newcounter{totalpagecounter}
\newcounter{totalfigurecounter}
\newcounter{totaltablecounter}
\newcounter{totalcitecounter}
\newcounter{totalpages}
\newcounter{totalfigures}
\newcounter{totaltables}
\newcounter{totalcites}
\makeatletter
\@input{\jobname.myaux}
\makeatother

\definecolor{white}{rgb}{1,1,1}
\definecolor{black}{rgb}{0,0,0}
\definecolor{red}{rgb}{1,0,0}
\definecolor{green}{rgb}{0,1,0}
\definecolor{blue}{rgb}{0,0,1}
\definecolor{cyan}{rgb}{0,1,1}
\definecolor{magenta}{rgb}{1,0,1}
\definecolor{yellow}{rgb}{1,1,0}
\definecolor{darkgreen}{rgb}{0,0.6,0}
\definecolor{darkyellow}{rgb}{0.8,0.8,0}
\definecolor{orange}{rgb}{1,0.5,0}
\definecolor{tuc}{RGB}{0,90,70}
\definecolor{tuclight}{RGB}{218,234,194}
\definecolor{tucorange}{RGB}{242,148,0}
\definecolor{tucbg}{RGB}{224,233,233}

\usepackage[
	colorlinks=true,                   
	urlcolor=blue,                     
	filecolor=blue,                    
	linkcolor=blue,                    
	citecolor=blue,                    
	breaklinks=true,                   
	backref=page,                      
	pagebackref=true,                  
	bookmarks=true,                    
	bookmarksopen=false,               
	bookmarksnumbered=true,            
	pdfauthor={Fabian Teichert},       
	pdfdisplaydoctitle=true,           
	pdfstartview=FitH,                 
	pdfpagelayout=TwoPageRight,        
]{hyperref}

\newif\ifsinglepaper\singlepaperfalse
\addto\captionsenglish{
	\renewcommand\bibname{References}
}

\renewcommand*{\backref}[1]{}
\renewcommand*{\backrefalt}[4]{%
	\ifsinglepaper\else%
		\ifcase #1
		\or (cited at p.~#2).
		\else (cited at pp.~#2).
		\fi%
	\fi
}
\usepackage{cite} 

\newcommand{\bstindent}{99}
\newcommand{\bstaddress}{}
\newcommand{\bstauthor}{}

\newcommand{\bstjournal}{}

\newcommand{\bstpublisher}{}

\newcommand{\bsttitle}{\itshape}
\newcommand{\bstvolume}{}
\newcommand{\bstyear}{}
\newcommand{\bbland}{and}

\newcommand{\bblnov}{November}

\newcommand\bibliographysection{\section}
\newcommand\bibliographysectionstyle{}
\newcommand\bibliographyitemsize{\normalsize}
\newcommand\bibliographyitemseparation{}
\makeatletter
\newcommand\bibcontentsline{\addcontentsline{toc}{section}{References}}
\renewenvironment{thebibliography}[1]{%
	\bibliographysection{\bibliographysectionstyle\bibname}
	\bibcontentsline%
	\renewcommand\rightmark{\bibname}
	\list{\@biblabel{\@arabic\c@enumiv}}{\settowidth\labelwidth{\@biblabel{#1}}%
		\leftmargin\labelwidth%
		\advance\leftmargin\labelsep%
		\@openbib@code%
		\usecounter{enumiv}%
		\let\p@enumiv\@empty%
		\renewcommand\theenumiv{\@arabic\c@enumiv}%
	}%
	\sloppy
	\clubpenalty4000
	\@clubpenalty \clubpenalty
	\widowpenalty4000%
	\sfcode`\.\@m%
}{%
	\def\@noitemerr{\@latex@warning{Empty `thebibliography' environment}}%
	\endlist%
}
\makeatother
\let\oldthebibliography\thebibliography
\renewcommand\thebibliography[1]{
	\bibliographyitemsize
	\oldthebibliography{#1}
	\bibliographyitemseparation
}

\let\oldtwocolumn\twocolumn
\let\oldonecolumn\onecolumn
\newif\iftwocolumn\twocolumntrue
\def\onecolumn{\twocolumnfalse}
\def\twocolumn{\twocolumntrue}
\newif\ifarticlestyle\articlestylefalse

\renewcommand{\title}[2]{%
	\setcounter{authors}{0}%
	\setcounter{addresses}{0}%
	\setcounter{keywords}{0}%
	\def\inserttitle{#1}%
	\def\articlelabel{#2}
}
\newcommand{\email}[1]{\def\insertemail{#1}}
\newcommand{\abstract}[1]{\def\insertabstract{#1}}
\def\insertjournal{}
\def\insertjournalshort{}
\def\insertdoi{}
\def\insertarxiv{}
\def\insertarxivshort{}
\newcommand{\journal}[4][nothing]{%
	\def\insertjournalshort{#2}%
	\if\relax\detokenize{#3}\relax%
		\def\insertjournal{}%
	\else%
		\def\tmpa{#1}%
		\def\tmpb{submitted}%
		\def\tmpc{accepted}%
		\ifx\tmpa\tmpb%
			\def\journalpre{Submitted to: }%
		\else%
			\ifx\tmpa\tmpc%
				\def\journalpre{Accepted in: }%
			\else%
				\def\journalpre{}%
			\fi%
		\fi%
		\if\relax\detokenize{#4}\relax\def\insertjournal{\journalpre#3}\else\def\insertjournal{\journalpre\href{#4}{#3}}\fi%
	\fi%
}
\newcommand{\doi}[1]{\if\relax\detokenize{#1}\relax\def\insertdoi{}\else\def\insertdoi{DOI: \href{http://dx.doi.org/#1}{#1}}\fi}
\newcommand{\arxiv}[2]{\if\relax\detokenize{#1}\relax\def\insertarxiv{}\def\insertarxivshort{}\else\def\insertarxiv{arXiv: \href{https://arxiv.org/abs/#1}{#1 [#2]}}\def\insertarxivshort{arXiv: #1}\fi}
\newcounter{authors}
\newcounter{addresses}
\newcounter{keywords}
\newcommand{\addauthor}[2]{\csdef{author\arabic{authors}}{#1}\csdef{authoraddress\arabic{authors}}{#2}\stepcounter{authors}}
\newcommand{\addaddress}[1]{\csdef{address\arabic{addresses}}{#1}\stepcounter{addresses}}
\newcommand{\addkeyword}[1]{\csdef{keyword\arabic{keywords}}{#1}\stepcounter{keywords}}

\newcounter{otherchapter}
\newcounter{normalchapter}
\newcounter{othercounter}
\newcounter{i}
\newcounter{j}
\makeatletter
\let\normalchapter\chapter

\renewcommand\chapter{%
	\@ifstar{%
		\normalchapter*%
	}{%
		\stepcounter{normalchapter}%
		\normalchapter%
	}%
}
\newcommand\otherchapter{%
	\protected@write\@auxout{}{\string\@writefile{lof}{\string\addvspace{10\string\p@}}}%
	\protected@write\@auxout{}{\string\@writefile{lot}{\string\addvspace{10\string\p@}}}%
	\scr@startsection{chapter}{1}{\z@}{0ex \@plus -0.2ex}{3.5ex \@plus 0.2ex}{\Large\bfseries}%
}
\newcommand\reftype{}
\newcommand\reflabel{}
\newcommand\refnumber{}
\newcommand\refshortnumber{}
\newcommand\reftext{}
\newcommand{\articletitlesub}{%
	\renewcommand\reftype{chapter}%
	\renewcommand\reflabel{section*.\arabic{othercounter}}%
	\renewcommand\refnumber{\Alph{otherchapter}}%
	\renewcommand\refshortnumber{}%
	\renewcommand\reftext{\inserttitle\ifx\insertjournalshort\empty\ (\insertarxivshort)\else\ (\insertjournalshort)\fi}%
	\pdfbookmark[0]{\reftext}{\reflabel}%
	\otherchapter*{\inserttitle}%
	\protected@write\@auxout{}{\string\@writefile{toc}{\string\contentsline {\reftype}{\string\numberline {\refnumber}\reftext}{\thepage}{\reflabel}}}%
	\Alabel{\articlelabel}%
	\noindent\textbf{%
		\large\csuse{author0}$^{\csuse{authoraddress0}}$%
		\forloop{i}{1}{\value{i} < \value{authors}}{%
			, \csuse{author\arabic{i}}$^{\csuse{authoraddress\arabic{i}}}$%
		}
	}\\[1em]
	\normalsize
	\setcounter{j}{0}
	\forloop{i}{0}{\value{i} < \value{addresses}}{%
		\stepcounter{j}
		$^{\arabic{j}}$\,\csuse{address\arabic{i}}
		\ifthenelse{\value{j}<\value{addresses}}{\\}{}
	}
	\ifx\insertemail\empty\\[1em]\else\\[0.5em]E-mail address: \insertemail\\[1em]\fi
	\textbf{Abstract:} \insertabstract
	\ifthenelse{\value{keywords}=0}{}{
		\\[1em]
		Keywords: \csuse{keyword0}%
			\forloop{i}{1}{\value{i} < \value{keywords}}{%
				; \csuse{keyword\arabic{i}}%
			}
	}
}
\newcommand{\articletitle}{%
	\setcounter{articlepage}{0}%
	\stepcounter{otherchapter}%
	\stepcounter{chapter}%
	\setcounter{section}{0}%
	\setcounter{subsection}{0}%
	\setcounter{subsubsection}{0}%
	\stepcounter{othercounter}%
	\iftwocolumn\oldtwocolumn[\articletitlesub\vspace{1.5em}]\else\oldonecolumn\articletitlesub\fi%
}%
\let\oldchapter\chapter
\let\oldsection\section
\let\oldsubsection\subsection
\let\oldsubsubsection\subsubsection
\newcommand\articlesectiondata[1]{%
	\renewcommand\reftype{section}%
	\renewcommand\reflabel{section.\arabic{chapter}.\arabic{section}}%
	\renewcommand\refnumber{\Alph{otherchapter}.\arabic{section}}%
	\renewcommand\refshortnumber{\arabic{section}}%
	\renewcommand\reftext{#1}%
}
\newcommand\articlesubsectiondata[1]{%
	\renewcommand\reftype{subsection}%
	\renewcommand\reflabel{subsection.\arabic{chapter}.\arabic{section}.\arabic{subsection}}%
	\renewcommand\refnumber{\Alph{otherchapter}.\arabic{section}.\arabic{subsection}}%
	\renewcommand\refshortnumber{\arabic{section}.\arabic{subsection}}%
	\renewcommand\reftext{#1}%
}
\newcommand\articlesectionnostar[1]{%
	\articlesectiondata{#1}%
	\pdfbookmark[1]{\reftext}{\reflabel}%
	\scr@startsection{section}{1}{\z@}{-3.5ex \@plus -1ex \@minus -0.2ex}{2.3ex \@plus 0.2ex}{\normalfont\bfseries}{#1}%
	\protected@write\@auxout{}{\string\@writefile{toc}{\string\contentsline {\reftype}{\string\numberline {\refnumber}#1}{\thepage}{\reflabel}}}%
}
\newcommand\articlesectionstar[1]{%
	\articlesectiondata{#1}%
	\scr@startsection{section}{1}{\z@}{-3.5ex \@plus -1ex \@minus -0.2ex}{2.3ex \@plus 0.2ex}{\normalfont\bfseries}*{#1}%
}
\newcommand\articlesubsectionnostar[1]{%
	\articlesubsectiondata{#1}%
	\pdfbookmark[2]{\reftext}{\reflabel}%
	\scr@startsection{subsection}{2}{\z@}{-3.5ex \@plus -1ex \@minus -0.2ex}{2.3ex \@plus 0.2ex}{\normalfont\bfseries}{#1}%
	\protected@write\@auxout{}{\string\@writefile{toc}{\string\contentsline {\reftype}{\string\numberline {\refnumber}#1}{\thepage}{\reflabel}}}%
}
\newcommand\articlesubsectionstar[1]{%
	\articlesubsectiondata{#1}%
	\scr@startsection{subsection}{2}{\z@}{-3.5ex \@plus -1ex \@minus -0.2ex}{2.3ex \@plus 0.2ex}{\normalfont\bfseries}*{#1}%
}
\newcommand\articlesection{\@ifstar{\stepcounter{othercounter}\articlesectionstar}{\articlesectionnostar}}
\newcommand\articlesubsection{\@ifstar{\stepcounter{othercounter}\articlesubsectionstar}{\articlesubsectionnostar}}
\renewcommand\chapter{\@ifstar{\stepcounter{othercounter}\oldchapter*}{\oldchapter}}
\renewcommand\section{\@ifstar{\stepcounter{othercounter}\oldsection*}{\oldsection}}
\renewcommand\subsection{\@ifstar{\stepcounter{othercounter}\oldsubsection*}{\oldsubsection}}
\renewcommand\subsubsection{\@ifstar{\stepcounter{othercounter}\oldsubsubsection*}{\oldsubsubsection}}
\newcommand\listof{}

\newcommand\articlefiguredata{%
	\renewcommand\listof{lof}%
	\renewcommand\reftype{figure}%
	\renewcommand\reflabel{figure.\arabic{chapter}.\arabic{figure}}%
	\renewcommand\refnumber{\Alph{otherchapter}.\arabic{figure}}%
	\renewcommand\refshortnumber{\arabic{figure}}%
}
\newcommand\articletabledata{%
	\renewcommand\listof{lot}%
	\renewcommand\reftype{table}%
	\renewcommand\reflabel{table.\arabic{chapter}.\arabic{table}}%
	\renewcommand\refnumber{\Alph{otherchapter}.\arabic{table}}%
	\renewcommand\refshortnumber{\arabic{table}}%
}
\renewenvironment{figure}{\articlefiguredata\begin{oldfigure}}{\end{oldfigure}} 
\renewenvironment{table}{\articletabledata\begin{oldtable}}{\end{oldtable}} 

\newenvironment{articlefigure*}{\articlefiguredata\begin{figure*}}{\end{figure*}}

\newenvironment{articletable*}{\articletabledata\begin{table*}}{\end{table*}}
\let\oldcaption\caption
\newcommand\Acaption[2][]{%
	\oldcaption[#1]{#2}%
	\renewcommand\reftext{#1}%
	\protected@write\@auxout{}{\string\@writefile{\listof}{\string\contentsline {\reftype}{\string\numberline {\refnumber}#1}{\thepage}{\reflabel}}}%
}
\renewcommand\caption[2][]{\ifarticlestyle\Acaption[#1]{#2}\else\oldcaption[#1]{#2}\fi}

\newcommand\botholdlabel[1]{\oldlabel{#1}\oldlabel{A#1}}
\newenvironment{articleequation}{\begin{equation}\renewcommand\label{\botholdlabel}}{\end{equation}} 
\newenvironment{articlesubequations}{\begin{subequations}\renewcommand\label{\botholdlabel}}{\end{subequations}} 
\newcommand\Aref[1]{\oldref{A#1}}
\newcommand\Alabel[1]{%
	\protected@write\@auxout{}{\string\newlabel{#1}{{\refnumber}{\thepage}{\reftext}{\reflabel}{}}}%
	\protected@write\@auxout{}{\string\newlabel{A#1}{{\refshortnumber}{\thepage}{\reftext}{\reflabel}{}}}%
}
\AtBeginDocument{%
	\let\oldref\ref%
	\let\oldlabel\label%
	\renewcommand\ref[1]{\ifarticlestyle\Aref{#1}\else\oldref{#1}\fi}%
	\renewcommand\label[1]{\ifarticlestyle\Alabel{#1}\else\oldlabel{#1}\fi}%
}
\makeatother

\newcommand\articlestyleheaderleft{%
	\ifx\insertarxiv\empty%
		\ifx\insertjournal\empty\linebreak\textnormal\insertdoi\else\linebreak\textnormal\insertjournal\fi%
	\else%
		\ifx\insertdoi\empty\linebreak\textnormal\insertjournal\else\textnormal\insertjournal\linebreak\textnormal\insertdoi\fi%
	\fi%
}
\newcommand\articlestyleheaderright{%
	\ifx\insertarxiv\empty%
		\ifx\insertjournal\empty\else\linebreak\textnormal\insertdoi\fi%
	\else%
		\linebreak\textnormal\insertarxiv%
	\fi%
}
\newcommand\articlestyleheadercenter{%
	\linebreak\textnormal\thechapter
}

\newcounter{articlepage}
\newcommand\articlepagemark{\arabic{articlepage}}

\newcommand\nocontentsline[3]{}
\let\oldaddcontentsline\addcontentsline
\newcommand\normalstyle{%
	\articlestylefalse%
	\KOMAoptions{fontsize=11pt}%
	\newgeometry{left=3cm,right=2.5cm,top=4cm,bottom=4cm}
	\setlength{\headheight}{26pt}
	\setlength{\headsep}{24pt}
	\setlength{\footskip}{30pt}
	\setlength{\TwoColumnWidth}{\textwidth}
	\setlength{\OneColumnWidth}{0.5\TwoColumnWidth-0.5\columnsep}
	\clearpairofpagestyles%
	\ihead{}%
	\chead{}%
	\ohead{\ifthispageodd{\textnormal\rightmark}{\textnormal\leftmark}}%
	\ifoot{}%
	\cfoot{}%
	\ofoot[\textnormal\pagemark]{\textnormal\pagemark}%
	\let\addcontentsline\oldaddcontentsline%
	\renewcommand{\thechapter}{\arabic{normalchapter}}%
	\renewcommand{\thesection}{\arabic{normalchapter}.\arabic{section}}%
	\renewcommand{\thesubsection}{\arabic{normalchapter}.\arabic{section}.\arabic{subsection}}%
	\renewcommand{\thesubsubsection}{\arabic{normalchapter}.\arabic{section}.\arabic{subsection}.\arabic{subsubsection}}%
	\renewcommand{\thefigure}{\arabic{normalchapter}.\arabic{figure}}%
	\renewcommand{\thetable}{\arabic{normalchapter}.\arabic{table}}%
	\renewcommand{\theequation}{\arabic{normalchapter}.\arabic{equation}}%
	\renewcommand\bibliographysection{\section*}%
	\renewcommand\bibcontentsline{\addcontentsline{toc}{section}{References}}
	\renewcommand\bibliographysectionstyle{}%
	\renewcommand\bibliographyitemsize{\normalsize}%
	\renewcommand\bibliographyitemseparation{%
		\setlength{\parskip}{0pt}%
		\setlength{\itemsep}{5pt plus 0.3ex}%
	}%
	\allowdisplaybreaks%
}
\newcommand\articlestyle{%
	\articlestyletrue%
	\KOMAoptions{fontsize=10pt}%
	\newgeometry{left=1.5cm,right=1.5cm,top=2.95cm,bottom=1.55cm}
	\setlength{\headheight}{24pt}
	\setlength{\headsep}{20pt}
	\setlength{\footskip}{1.2cm}
	\setlength{\TwoColumnWidth}{\textwidth}
	\setlength{\OneColumnWidth}{0.5\TwoColumnWidth-0.5\columnsep}
	\clearpairofpagestyles%
	\ihead{\ifthispageodd{\articlestyleheaderleft}{\articlestyleheaderright}}%
	\chead{\ifsinglepaper\else\articlestyleheadercenter\fi}%
	\ohead{\ifthispageodd{\articlestyleheaderright}{\articlestyleheaderleft}}%
	\ifoot{}%
	\cfoot{\ifsinglepaper\textnormal\pagemark\else\stepcounter{articlepage}\textnormal{\thechapter-\articlepagemark}\fi}%
	\ofoot{\ifsinglepaper\else\textnormal\pagemark\fi}%
	\let\addcontentsline\nocontentsline%
	\renewcommand{\thechapter}{\Alph{otherchapter}}%
	\renewcommand{\thesection}{\arabic{section}}%
	\renewcommand{\thesubsection}{\arabic{section}.\arabic{subsection}}%
	\renewcommand{\thesubsubsection}{\arabic{section}.\arabic{subsection}.\arabic{subsubsection}}%
	\renewcommand{\thefigure}{\arabic{figure}}%
	\renewcommand{\thetable}{\arabic{table}}%
	\renewcommand{\theequation}{\arabic{equation}}%
	\renewcommand\bibliographysection{\articlesection*}%
	\renewcommand\bibcontentsline{\oldaddcontentsline{toc}{section}{References}}
	\renewcommand\bibliographysectionstyle{\normalsize}%
	\renewcommand\bibliographyitemsize{\small}%
	\renewcommand\bibliographyitemseparation{%
		\setlength{\parskip}{0pt}%
		\setlength{\itemsep}{0pt plus 0.3ex}%
	}%
	\interdisplaylinepenalty=10000%
}
\normalstyle

\renewcommand{\hbar}{\mathchar'26\mkern-9mu \mathrm{h}}
\newcommand{\one}{\mathcal{I}}
\newcommand{\hamilton}{\mathcal{H}}

\newcommand{\coupling}{\tau}

\newcommand{\green}{\mathcal{G}}

\newcommand{\transmission}{\mathcal{T}}

\newcommand{\imag}{\text{i}}

\newcommand{\trace}[1]{\text{Tr}\!\left[#1\right]} 
\newcommand{\order}[1]{\mathcal{O}\!\left(#1\right)}

\singlepapertrue
\begin{document}

\raggedbottom

\frontmatter
\clearpairofpagestyles
\ofoot[\textnormal\pagemark]{\textnormal\pagemark}
\KOMAoptions{headsepline=false}

\mainmatter
\KOMAoptions{headsepline=true}

\normalstyle

\articlestyle

\renewcommand{\one}{\mathcal{I}}
\renewcommand{\hamilton}{\mathcal{H}}
\renewcommand{\coupling}{\tau}
\renewcommand{\green}{\mathcal{G}}
\renewcommand{\transmission}{\mathcal{T}}
\renewcommand{\imag}{\text{i}}
\renewcommand{\order}[1]{\mathcal{O}\!\left(#1\right)}
\renewcommand{\trace}[1]{\text{Tr}\!\left(#1\right)} 
\newcommand{\intd}[4]{\int\limits_{#1}^{#2}#3\text{d}#4} 

\onecolumn 

\title{Improved recursive Green's function formalism for quasi one\hyp{}dimensional systems with realistic defects}{JCP1}

\addauthor{Fabian Teichert}{1,3,4}
\addauthor{Andreas Zienert}{2}
\addauthor{J\"org Schuster}{3,4}
\addauthor{Michael Schreiber}{1,4}

\addaddress{Institute of Physics, Technische Universit\"at Chemnitz, 09107 Chemnitz, Germany}
\addaddress{Center for Microtechnologies, Technische Universit\"at Chemnitz, 09107 Chemnitz, Germany}
\addaddress{Fraunhofer Institute for Electronic Nano Systems (ENAS), 09126 Chemnitz, Germany}
\addaddress{Dresden Center for Computational Materials Science (DCMS), TU Dresden, 01062 Dresden, Germany}

\email{fabian.teichert@physik.tu-chemnitz.de}

\abstract{
We derive an improved version of the recursive Green's function formalism (RGF), which is a standard tool in the quantum transport theory.
We consider the case of disordered quasi one-dimensional materials where the disorder is applied in form of randomly distributed realistic defects, leading to partly periodic Hamiltonian matrices.
The algorithm accelerates the common RGF in the recursive decimation scheme, using the iteration steps of the renormalization decimation algorithm.
This leads to a smaller effective system, which is treated using the common forward iteration scheme.
The computational complexity scales linearly with the number of defects, instead of linearly with the total system length for the conventional approach.
We show that the scaling of the calculation time of the Green's function depends on the defect density of a random test system.
Furthermore, we discuss the calculation time and the memory requirement of the whole transport formalism applied to defective carbon nanotubes.
}

\addkeyword{recursive Green's function formalism (RGF)}
\addkeyword{renormalization decimation algorithm (RDA)}
\addkeyword{electronic transport}
\addkeyword{carbon nanotube (CNT)}
\addkeyword{defect}

\journal{J. Comput. Phys. 334 (2017), 607--619}{Journal of Computational Physics 334 (2017), 607--619}{http://www.sciencedirect.com/science/article/pii/S0021999117300347} 
\doi{10.1016/j.jcp.2017.01.024} 
\arxiv{1705.02178}{physics.comp-ph} 

\articletitle

\articlesection{Introduction}\label{JCP1:Introduction}

In the last decades, simulation techniques became more and more important for a huge diversity of topics in theoretical physics. Mechanical, optical, and electronic properties of new materials have been studied, often driven by developments in semiconductor technologies. Simultaneously, the miniaturization process shifts the dimensions of microelectronic devices into the mesoscopic range, which cannot be treated classically. Simulation methods have to deal with a large complexity concerning quantum mechanical effects, different materials and their interplay, and multiple length scales.

Electronic transport properties of arbitrary devices, which are connected to electrodes, can be described by electron scattering and a resulting transmission probability, using the Landauer-B\"uttiker formalism~\cite{PhysRevB.31.6207}. In combination with quantum transport theory~\cite{Datta2005} and an underlying electron structure theory like the tight binding method, density functional theory~\cite{BrazJPhys.36.1318}, or hybrid methods, this allows one to calculate currents and conductances through devices precisely via the Green's function of the system using its given Hamiltonian. This approach needs a computationally expensive inversion, whose complexity scales with the third power of the number of involved atoms. However, the recursive Green's function formalism (RGF)~\cite{JPhysCSolidStatePhys.5.2845, CompPhysCommun.20.11, JPhysCSolidStatePhys.14.235, ZPhysBCondMat.59.385} provides a very efficient, linearly scaling algorithm by dividing the system into layers. It can be applied to arbitrary systems which have a sparse Hamiltonian matrix, as is the case with all short-range interactions in real space. Although this is a common method nowadays, further technical improvements of the algorithm are necessary, because the numerical complexity still rapidly reaches the limits of today's computer resources for large and complex systems or when different physical effects (i.e. levels of theory) are considered.

It is still difficult to apply the RGF in an optimal way to arbitrary systems. For difficult geometries the question how to arrange the RGF layers or how to get the optimal block-tridiagonal matrix form has to be considered. Different solutions have been found, e.g. a pivoting-like bandwidth minimization method in combination with a block-tridiagonalization algorithm~\cite{JComputPhys.228.8548}, the reverse Cuthill-McKee algorithm for the block-tridiagonalization of connected graphs~\cite{PhysRevB.84.155401, ACM69.157}, the Knitting algorithm, which uses the Dyson equation and the resulting decimation scheme to build up the system site by site~\cite{PhysRevB.77.115119}, or accelerations by using singular value decompositions~\cite{PhysRevE.90.013306}. Besides the RGF, also other methods are often utilized. The inversion can be done directly, using fast LU-decomposition algorithms with nested dissection~\cite{SIAMJNumerAnal.10.345}. The simple basic idea, divide and conquer, leads to optimal scaling in the limit of large systems. But ``large'' in this context is until now still larger than systems which are treatable within acceptable time. The existing nested dissection method can be further enhanced by using width-one separators instead of width-two separators, which reduces calculation time drastically~\cite{JComputPhys.242.915}. LU-decomposition methods also allow to calculate selected elements of an inverse matrix very fast. This is especially interesting for transport problems, because only a few elements of the Green's function are necessary. Based on this, fast recurrence formulas can be found similar to the RGF but based on the LU-decomposition~\cite{JComputPhys.228.5020, EuroPar2013.533}. Also combinations with the RGF like the over-bridging boundary-matching method in combination with the shifted conjugate-orthogonal conjugate-gradient method~\cite{PhysRevB.67.195315, PhysRevB.73.165108, PhysRevB.86.195406, PhysRevE.91.063305, JResNBS.49.409} are promising.

In the present paper we describe new methodical improvements to speed up quantum transport calculations of certain quasi one-dimensional systems. In the sections \ref{JCP1:QTT} and \ref{JCP1:RGF} we shortly explain the most important formulas of (equilibrium) quantum transport theory and common solutions for quasi one-dimensional systems using the RGF. Based on this, in section \ref{JCP1:RGF-RDA}, we discuss an advanced solution for special cases and develop a further improvement of the RGF. We consider the case of disordered quasi one-dimensional systems where the disorder is caused by randomly positioned realistic defects in an otherwise ideal system. In contrast to Anderson-like disorder~\cite{PhysRev.109.1492, PhysRevLett.47.1546, PhysRevLett.42.673}, which randomizes the Hamiltonian matrix of the whole system, this system consists of repeated ideal parts which are interrupted by defective parts. That means that the Hamiltonian matrix has many blocks which are equal and only a few ones which are different. For such partly periodic cases the renormalization decimation algorithm (RDA)~\cite{JPhysFMetPhys.14.1205, JPhysFMetPhys.15.851}, which calculates the Green's function of the electrodes within an iteration process, allows us to treat the ideal parts much faster than the pure RGF can do. We discuss the calculation time and memory requirement for a simple random test system and especially for carbon nanotubes (CNTs) with divacancy defects.

CNTs are prominent examples for future electronic devices, because of their excellent electronic properties. But many things can negatively affect them. CNTs have to be integrated into conventional devices and thus are contacted to metals. The resulting interplay between these surrounding metals and the CNT leads to unintended reductions of the current and the conductance~\cite{Nanotechnology.25.425203, MicroEng.137.124, PhysRevB.91.165404}. Also defects have a huge impact. Several publications concerning the influence of vacancies~\cite{JPhysCondMat.20.294214, JPhysCondMat.20.304211, PhysRevLett.95.266801, JPhysCondMat.26.045303, JPhysChemC.116.1179}, substitutional atoms~\cite{SolidStateCommun.149.874}, functionalizations~\cite{PhysStatSolB.247.2962, NanoRes.3.288, NanoLett.9.940}, and Anderson disorder~\cite{PhysRevB.58.4882} showed that the ballistic transport is driven into the strong localization regime~\cite{PhysRevB.64.045409}. But mostly this has been done exemplary, as extensive calculations of CNT ensembles are necessary. We contributed to this task with a comprehensive study of the electronic properties of CNTs under the influence of randomly distributed monovacancies and divacancies. We varied the CNT length, the CNT diameter, the defect type, the defect probability, and the temperature, and we discussed the resulting simple analytical dependencies~\cite{NJPhys.16.123026}.

\articlesection{Quantum transport theory}\label{JCP1:QTT}

\begin{figure}[tb]
	\includegraphics{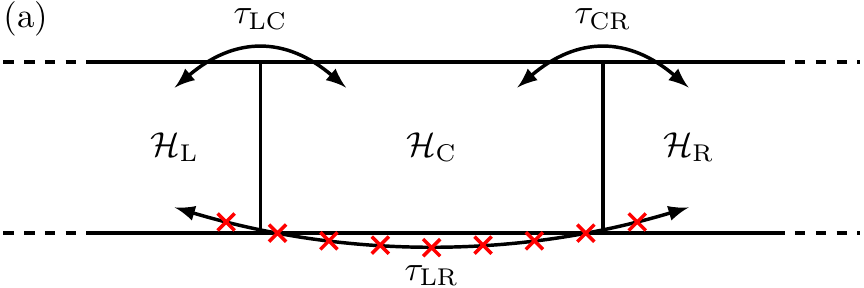}\hfill
	\includegraphics{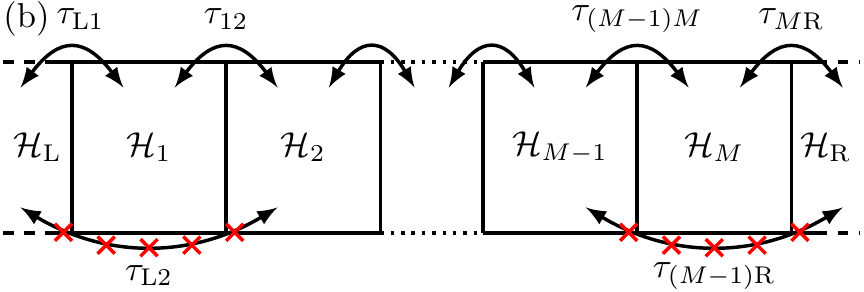}
	\caption[Scheme of a device system]{(Color online.) Scheme of the device system used in the transport formalism~\cite{NJPhys.16.123026}. The Hamiltonian matrices of the subsystems are denoted by $\hamilton$ and the coupling matrices between the subsystems by $\coupling$. Crossed-out couplings are not taken into account in our approach. (a) The two electrodes (L and R) are ideal, semi-infinite parts. The central region (C) includes the defective part. (b) The central region is divided into $N$ subsystems to which the RGF is applied.}
	\label{JCP1:fig:transport:scheme}
\end{figure}

In this section we give a brief introduction into (equilibrium) quantum transport theory and the resulting numerical challenges. Quantum transport theory describes transport properties like transmission and conductance (but also electronic structure properties like density of states and electron density) of open systems~\cite{Datta2005}. ``Open'' means that the system of interest (conductor) is connected to comparably large leads (that can be considered as reservoirs where electrons are always at their respective equilibrium), from/to which electrons are allowed to enter/leave. For this, the system is described as shown in figure \ref{JCP1:fig:transport:scheme}a. A finite central region C, which contains the total non-periodic part of the quasi one-dimensional system, is connected to two half-infinite and periodic electrodes (left L and right R), which act as reservoirs providing electrons or holes~\cite{Datta2005}. We call this a device configuration. The corresponding Schr\"odinger equation within an orthonormal basis\footnote{The Schr\"odinger equation within a non-orthogonal basis can be obtained by substituting $E$ by $E\mathcal{S}$. $\mathcal{S}$ is the overlap matrix, which has the same device block structure as the Hamiltonian matrix. This substitution can be applied to the subsequent equations as well.} reads
\begin{articleequation}
	\begin{pmatrix}
		\hamilton_\text{L} & \coupling_\text{LC} & 0 \\
		\coupling_\text{CL} & \hamilton_\text{C} & \coupling_\text{CR} \\
		0 & \coupling_\text{RC} & \hamilton_\text{R}
	\end{pmatrix}\varPsi = E\varPsi \qquad .\label{JCP1:eqn:SGL}
\end{articleequation}%
$\hamilton_\text{L/C/R}$ denote the Hamiltonian matrices of the parts L, C, and R. $\coupling$ denote the coupling matrices between two of these parts. If region C is long enough, the direct interaction of L and R can be neglected, $\coupling_\text{LR}=0$. Note that this is a matrix eigenvalue equation of infinite dimension, which in this form is numerically not treatable. However, it can be reduced to an effective matrix eigenvalue equation of finite dimension by treating the electrodes separately and including their influence as a self-energy correction.

We define the advanced Green's function matrix of the central region
\begin{articleequation}
	\green_\text{C}\!\left(E\right) = \lim_{\eta\rightarrow0^+} \left[ (E+\imag\eta)\one - \hamilton_\text{C} - \varSigma_\text{L} - \varSigma_\text{R} \right]^{-1} \qquad .\label{JCP1:eqn:Green}
\end{articleequation}%
$\varSigma_\text{L}=\coupling_\text{CL}\green_\text{L}\coupling_\text{LC}$ and $\varSigma_\text{R}=\coupling_\text{CR}\green_\text{R}\coupling_\text{RC}$ are self-energy matrices, which lead to an energy-dependent shift of the electronic states due to the electrode coupling. $\green_\text{L/R}$ are the advanced surface Green's functions of the electrodes, which can be calculated with the RDA (see section \ref{JCP1:RDA}). $\one$ is the identity matrix of appropriate dimension. The transmission spectrum $\transmission(E)$ of the device configuration can be calculated with
\begin{articleequation}
	\transmission(E) = \trace{\varGamma_\text{R} \green_\text{C} \varGamma_\text{L} \green_\text{C}^\dagger} \qquad .\label{JCP1:eqn:T(E)}
\end{articleequation}%
Therein, $\varGamma_\text{L/R}=\imag\left(\varSigma_\text{L/R}-\varSigma_\text{L/R}^\dagger\right)$ are broadening matrices, which lead to an energy-dependent broadening of the electronic states due to the electrode coupling. Finally the conductance is given by the Landauer-B\"uttiker formalism~\cite{PhysRevB.31.6207}
\begin{articleequation}
	G = -\text{G}_0\intd{-\infty}{\infty}{\transmission(E)f'(E)}{E} \qquad\text{with}\qquad \text{G}_0=\frac{2\text{e}^2}{\text{h}} \qquad\text{and}\qquad f(E)=\frac{1}{1+\text{exp}\left(\frac{E-E_\text{F}}{\text{k}_\text{B}T}\right)} \qquad .\label{JCP1:eqn:G}
\end{articleequation}%
$E_\text{F}$ is the Fermi energy.

In summary, calculating the transmission spectrum involves two main tasks: the electrode calculations (self-energies), in which infinite but periodic problems must be solved, and the inversion problem of the central region (which can be very large). In the range of mesoscopic systems with hundreds of thousands of atoms, the latter is a time and memory consuming process (the calculation complexity scales as $\mathcal{O}\!\left[(\dim\hamilton_\text{C})^3\right]$). However, we can take advantage of the block-tridiagonal shape of $\hamilton_\text{C}$. If region C is much longer than the interaction distance, it can be divided into $N$ subsystems with Hamiltonian matrices $\hamilton_i$ and coupling matrices $\coupling_{ij}$, as shown in figure~\ref{JCP1:fig:transport:scheme}b. As non-neighboring cells are not interacting, the corresponding coupling matrices are zero and the transmission spectrum can be calculated with the simplified formula
\begin{articleequation}
	\transmission(E) = \trace{\varGamma'_\text{R} \green_{N1} \varGamma'_\text{L} \green_{N1}^\dagger} \qquad .\label{JCP1:eqn:T(E):RGF}
\end{articleequation}%
$\green_{N1}$ is the lower left matrix block of $\green_\text{C}$ and $\varGamma'_\text{L}$ ($\varGamma'_\text{R}$) is the upper left (lower right) matrix block of $\varGamma_\text{L}$ ($\varGamma_\text{R}$).

\articlesection{Recursive Green's function formalisms}\label{JCP1:RGF}

The RGF~\cite{JPhysCSolidStatePhys.14.235, ZPhysBCondMat.59.385} is a method for calculating $\green_{N1}$. There exist different implementations of the RGF, which are widely used for mesoscopic systems with hundreds of thousands of atoms. In the following, we discuss the forward iteration scheme (FIS), the recursive decimation scheme (RDS), and the RDA~\cite{JPhysFMetPhys.14.1205, JPhysFMetPhys.15.851}. The latter is a convergent iterative algorithm for calculating the surface Green's functions of periodic systems.

\articlesubsection{Forward iteration scheme}\label{JCP1:FIS}

\begin{figure}[tb]
	\includegraphics{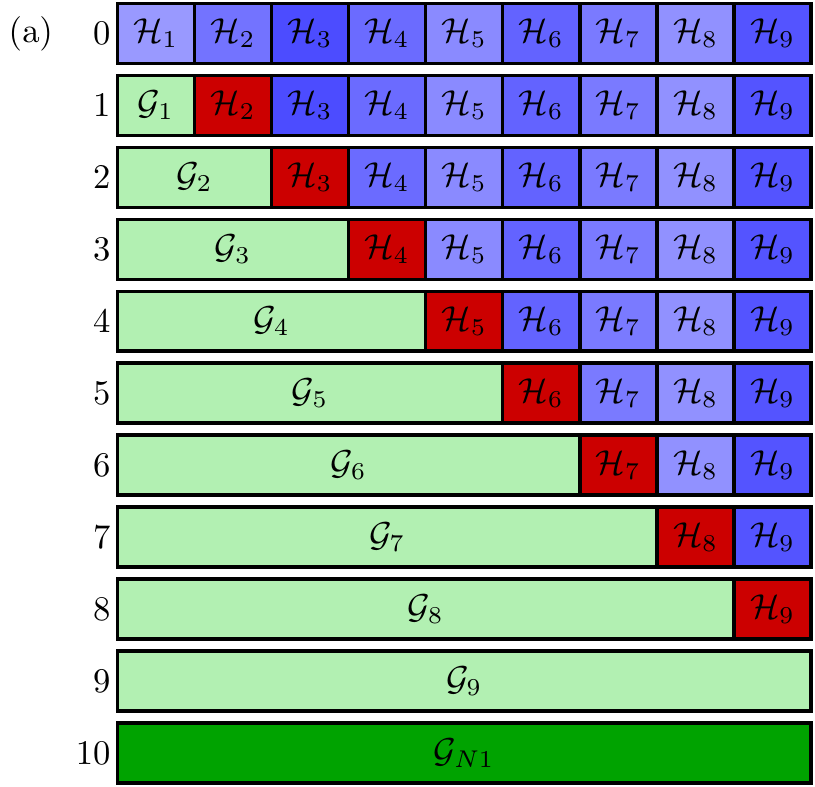}\hfill
	\includegraphics{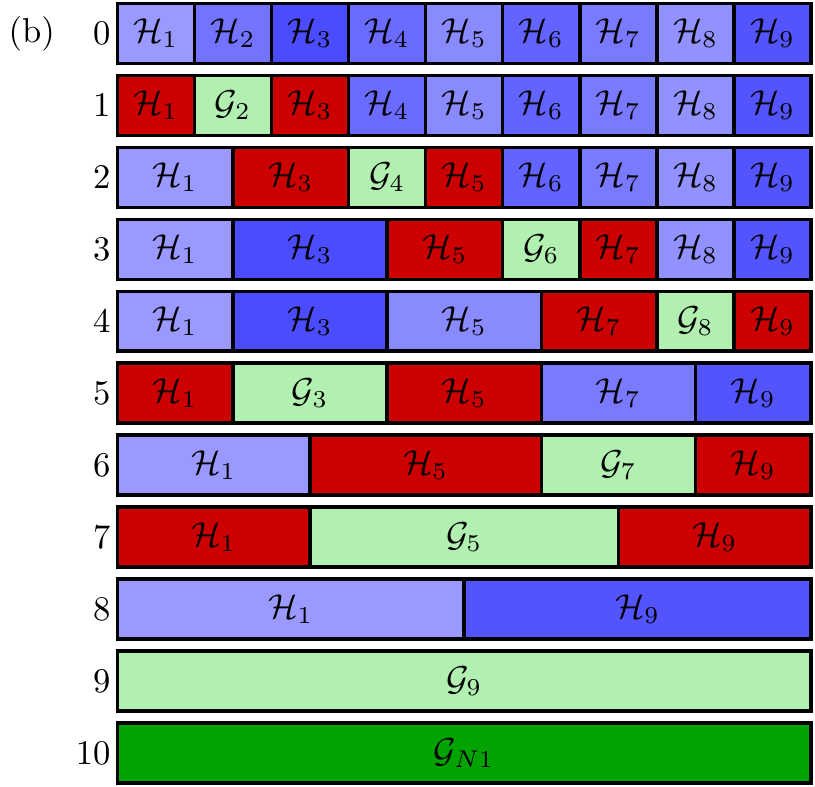}
	\caption[Sketch of the RGF-FIS and the RGF-RDS]{(Color online.) (a) Sketch of the RGF-FIS. (b) Sketch of the RGF-RDS. From top to bottom: Hamiltonian matrices and Green's matrices when applying equations (\ref{JCP1:eqn:RGF-FIS}) or (\ref{JCP1:eqn:RGF-RDS}). The numbers at the left of each line denote the iteration step. Different blue blocks denote different Hamiltonian matrices (defect cells). Green blocks denote the calculation of the Green's matrix of the actual iteration step. The dark green block denotes the final step. Red blocks denote Hamiltonian matrices which are coupled to the highlighted Green's matrix. They get a self-energy correction.}
	\label{JCP1:fig:RGF}
\end{figure}

The FIS is based on the inversion of a $2\times 2$ block matrix where only the lower left block of the inverse matrix is of interest for transport calculations.
\begin{align}
	\begin{pmatrix}A_{11}&A_{12}\\A_{21}&A_{22}\end{pmatrix}^{-1} = \begin{pmatrix}B_{11}&B_{12}\\B_{21}&B_{22}\end{pmatrix} \quad\Rightarrow\quad B_{21} = -\tilde{A}_{22}^{-1}A_{21}A_{11}^{-1} \quad\text{with}\quad \tilde{A}_{22}=A_{22}-A_{21}A_{11}^{-1}A_{12}
\end{align}%
This can be repeated by dividing $A_{22}$ into $2\times 2$ blocks and so on, leading to a matrix version of the Gau\ss-Jordan elimination. The resulting recursion formula can easily be transferred to the calculation of $\green_{N1}$. Additionally, noticing the self-energy corrections in the first and the last subsystem, we obtain
\begin{articlesubequations}
	\label{JCP1:eqn:RGF-FIS}
	\begin{align}
		\green_1 &= \lim_{\eta\rightarrow0^+} \left[ (E+\imag\eta)\one - \hamilton_1 - \varSigma_\text{L} \right]^{-1} \qquad ,\\
		\green_i &= \lim_{\eta\rightarrow0^+} \left[ (E+\imag\eta)\one - \hamilton_i - \coupling_{i(i-1)}\green_{i-1}\coupling_{(i-1)i} \right]^{-1} \quad\text{for}~2\leq i\leq N-1 \qquad ,\\
		\green_N &= \lim_{\eta\rightarrow0^+} \left[ (E+\imag\eta)\one - \hamilton_N - \coupling_{N(N-1)}\green_{N-1}\coupling_{(N-1)N} - \varSigma_\text{R} \right]^{-1} \qquad ,\\
		\mathcal{P}_{(i+1)i} &= \coupling_{(i+1)i}\green_i \quad\text{for}~1\leq i\leq N-1 \qquad ,\\
		\green_{N1} &= \green_N\mathcal{P}_{N(N-1)}\mathcal{P}_{(N-1)(N-2)}\cdots\mathcal{P}_{32}\mathcal{P}_{21} \qquad .
	\end{align}%
\end{articlesubequations}%
The computational complexity of this algorithm scales as $\order{N[\dim\hamilton_\text{i}]^3}$, which is a factor $N^2$ better than a direct inversion of the Hamiltonian matrix. A sketch of the RGF-FIS is shown in figure \ref{JCP1:fig:RGF}a.

\articlesubsection{Recursive decimation scheme}\label{JCP1:RDS}

The basic idea of the RDS is similar to the previous scheme. The inversion of a $2\times 2$ block matrix where the second row and second column of the inverse matrix are irrelevant, can be reduced to an effective inversion problem of the first block.
\begin{align}
	\begin{pmatrix}A_{11}&A_{12}\\A_{21}&A_{22}\end{pmatrix} \begin{pmatrix}B_{11}&B_{12}\\B_{21}&B_{22}\end{pmatrix} = \begin{pmatrix}\one&0\\0&\one\end{pmatrix} \quad\Rightarrow\quad \tilde{A}_{11}B_{11} = \one \quad\text{with}\quad \tilde{A}_{11}=A_{11}-A_{12}A_{22}^{-1}A_{21}
\end{align}%
Concerning the $N$ parts of the Hamiltonian matrix $\hamilton_\text{C}$ and the relevant Green's matrix block $\green_{N1}$, the subsystems $\hamilton_2,\ldots,\hamilton_{N-1}$ can be decimated using this scheme.
\begin{articlesubequations}
	\label{JCP1:eqn:RGF-RDS}
	\begin{flalign}
		\omit\rlap{\text{For $k\in\{1,2,\ldots,\lceil\log_2(N-1)\rceil\}$:}} &&\\ 
		\omit\rlap{\hspace{3em}\text{For $l\in\left\{0,1,\ldots,\left\lceil\frac{N-1-2^{k-1}}{2^k}-1\right\rceil\right\}$:}} &&\\ 
		\hspace{6em} i &:= 1+2^{k-1}+l\cdot 2^k &\\
		\hspace{6em} j_\text{L} &:= 1+l\cdot 2^k &\\
		\hspace{6em} j_\text{R} &:= 1+(l+1)\cdot 2^k &\\
		\hspace{6em} \green_i &:= \lim_{\eta\rightarrow0^+} \left[ (E+\imag\eta)\one - \hamilton_i \right]^{-1} &\\
		\hspace{6em} \hamilton_j &:= \hamilton_j + \coupling_{ji}\green_i\coupling_{ij} \quad\forall~j\in\left\{j_\text{L},j_\text{R}\right\} &\\
		\hspace{6em} \coupling_{j_\text{L}j_\text{R}} &:= \coupling_{j_\text{L}i}\green_i\coupling_{ij_\text{R}} &\\
		\hspace{6em} \coupling_{j_\text{R}j_\text{L}} &:= \coupling_{j_\text{R}i}\green_i\coupling_{ij_\text{L}} &
	\end{flalign}
\end{articlesubequations}%
Here, $\lceil\cdot\rceil$ is the ceiling function, $i$ is the actual cell being decimated, and $j_\text{L/R}$ is the next left/right cell that has not been decimated so far. In a last step, $\green_{N1}$ can be calculated using equations (\ref{JCP1:eqn:RGF-FIS}) for a two part system consisting of $\hamilton_1$ and $\hamilton_N$. The comparison of equations (\ref{JCP1:eqn:RGF-FIS}) with equations (\ref{JCP1:eqn:RGF-RDS}) shows that the FIS is preferable as it needs $N$ inversions and $3N-3$ multiplications, in contrast to the RDS with $N$ inversions and $6N-9$ multiplications. A sketch of the RGF-RDS is shown in figure \ref{JCP1:fig:RGF}b.

\articlesubsection{Renormalization decimation algorithm}\label{JCP1:RDA}

\begin{figure}[tb]
	\centering
	\includegraphics{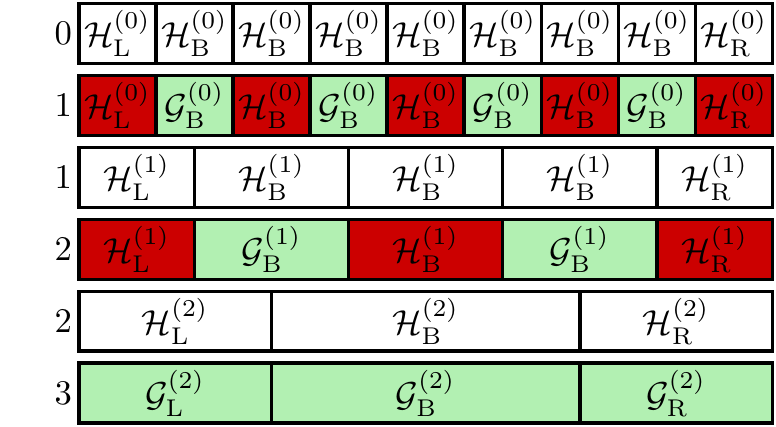}
	\caption[Sketch of the RDA]{(Color online.) Sketch of the RDA. From top to bottom: Hamiltonian matrices and Green's matrices when applying equations (\ref{JCP1:eqn:RDA}). The numbers at the left of each line denote the iteration step. White blocks denote same Hamiltonian matrices (ideal cells). Green blocks denote the calculation of the Green's matrix of the actual iteration step. Red blocks denote Hamiltonian matrices which are coupled to the highlighted Green's matrices. They get a self-energy correction.}
	\label{JCP1:fig:RDA}
\end{figure}

The RDA is a very fast version of the RGF-RDS for calculating the left (L), right (R), and bulk (B) Green's function of a (half-)infinite periodic system, e.g. the electrodes of a device. Due to the periodicity, this can be done in a convergent iterative process of equations (\ref{JCP1:eqn:RGF-RDS}) with $N\rightarrow\infty$. For simplicity, let us consider a system with $2^k+1$ parts. As all subsystems are equal, non-neighboring cells can be decimated in one step. Decimating every second cell, beginning with the second, is most efficient. After this step all remaining subsystems are again equal, except the first and the last one, which will not be decimated. This scheme can be used for all following steps, taking into account that the first and the last cell differ. In the last step, three effective cells remain corresponding to the left, right, and bulk Green's function. The overall RDA reads
\begin{articlesubequations}
	\label{JCP1:eqn:RDA}
	\begin{align}
		\green_\text{L/B/R}^{(i)} &= \lim_{\eta\rightarrow0^+} \left[ (E+\imag\eta)\one - \hamilton_\text{L/B/R}^{(i)} \right]^{-1}  \quad\text{with}\quad \hamilton_\text{L/B/R}^{(0)} = \hamilton \qquad,\\
		\alpha^{(i+1)} &= \alpha^{(i)}\green^{(i)}_\text{B}\alpha^{(i)} \quad\text{with}\quad \alpha^{(0)} = \coupling \qquad,\\
		\beta^{(i+1)} &= \beta^{(i)}\green^{(i)}_\text{B}\beta^{(i)} \quad\text{with}\quad \beta^{(0)} = \coupling^\dagger \qquad,\\
		\hamilton_\text{B}^{(i+1)} &= \hamilton_\text{B}^{(i)} + \alpha^{(i)}\green_\text{B}^{(i)}\beta^{(i)} + \beta^{(i)}\green_\text{B}^{(i)}\alpha^{(i)} \qquad,\\
		\hamilton_\text{L}^{(i+1)} &= \hamilton_\text{L}^{(i)} + \alpha^{(i)}\green_\text{B}^{(i)}\beta^{(i)} \qquad,\\
		\hamilton_\text{R}^{(i+1)} &= \hamilton_\text{R}^{(i)} + \beta^{(i)}\green_\text{B}^{(i)}\alpha^{(i)} \qquad.
	\end{align}%
\end{articlesubequations}%
$\coupling$ is the coupling matrix between the equal subsystems, $\alpha^{(i)}$ and $\beta^{(i)}$ are effective coupling matrices. The iteration can be considered converged if $\left\|\alpha\right\|+\left\|\beta\right\|$ falls below some threshold. A finite system with $2^k+1$ parts needs $k-1$ inversions and $6k-6$ multiplications. Considering a system with $N$ parts, $\log_2(N+1)-1$ inversions and $6\log_2(N+1)-6$ multiplications are needed. A sketch of the RDA is shown in figure \ref{JCP1:fig:RDA}.

\articlesection{Improved RGF+RDA}\label{JCP1:RGF-RDA}

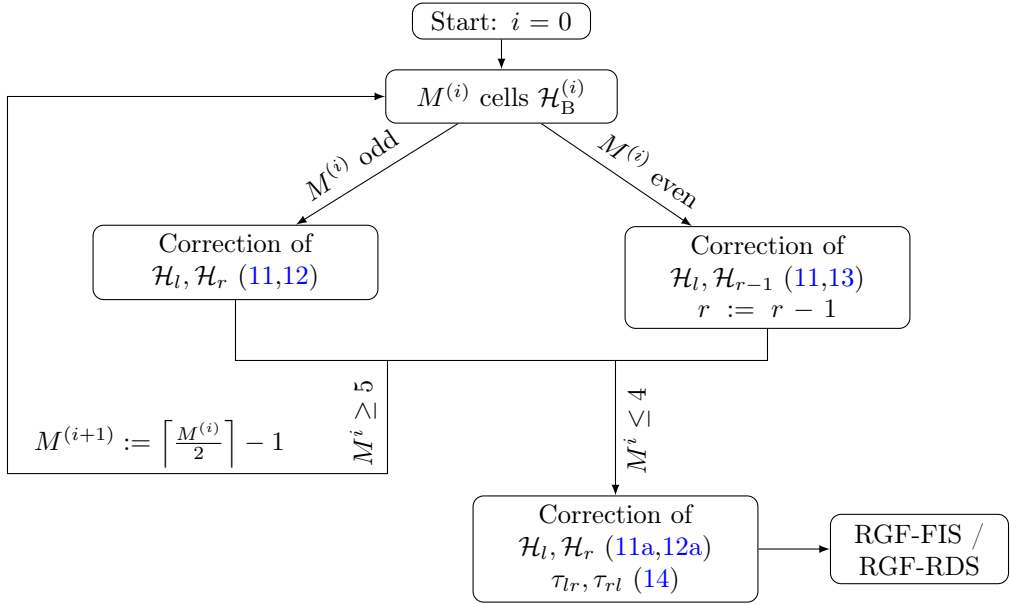
\begin{figure}[tb]
	\centering
	\begin{tikzpicture}
		\node[draw=black, rounded corners, text width=6em, text centered] (a) at (0,6.5) {Start: $i=0$};
		\node[draw=black, rounded corners, text width=8em, text centered] (b) at (0,5.5) {$M^{(i)}$ cells $\hamilton_\text{B}^{(i)}$};
		\node[draw=black, rounded corners, text width=10em, text centered] (ca) at (-3.5,3.3) {Correction of\\$\hamilton_l,\hamilton_r$ (\ref{JCP1:eqn:RGF-RDA:left},\ref{JCP1:eqn:RGF-RDA:right:odd})};
		\node[draw=black, rounded corners, text width=10em, text centered] (cb) at (3.5,3.1) {Correction of\\$\hamilton_l,\hamilton_{r-1}$ (\ref{JCP1:eqn:RGF-RDA:left},\ref{JCP1:eqn:RGF-RDA:right:even})\\$r:=r-1$};
		\node[draw=black, rounded corners, text width=10em, text centered] (d) at (1.5,-0.5) {Correction of\\$\hamilton_l,\hamilton_r$ (\oldref{JCP1:eqn:RGF-RDA:left:H},\oldref{JCP1:eqn:RGF-RDA:right:odd:H})\\$\coupling_{lr},\coupling_{rl}$ (\ref{JCP1:eqn:RGF-RDA:final})};
		\node[draw=black, rounded corners, text width=6em, text centered] (e) at (5.5,-0.5) {RGF-FIS / RGF-RDS};
		\draw[-latex] (a) -> (b);
		\draw[-latex] (b) -> (ca);
		\draw[-latex] (b) -> (cb);
		\draw (ca) -- (-3.5,2) -- (3.5,2) -- (cb);
		\draw[-latex] (-1.5,2) -> (-1.5,0.5) -> (-6.5,0.5) -> (-6.5,5.5) -> (b);
		\draw[-latex] (1.5,2) -> (d);
		\draw[-latex] (d) -> (e);
		\node at (-2.5,4.1) {\begin{rotate}{34}$M^{(i)}$ odd\end{rotate}};
		\node at (1.25,4.85) {\begin{rotate}{-34}$M^{(i)}$ even\end{rotate}};
		\node at (-1.7,0.6) {\begin{rotate}{90}$M^{i}\geq 5$\end{rotate}};
		\node at (1.9,0.5) {\begin{rotate}{90}$M^{i}\leq 4$\end{rotate}};
		\node at (-4.5,0.9) {$M^{(i+1)}:=\left\lceil\tfrac{M^{(i)}}{2}\right\rceil-1$};
	\end{tikzpicture}
	\caption[Scheme of the RGF+RDA]{Scheme of the improved RGF+RDA. A periodic part of the system consisting of $M^{(i)}$ ideal cells is treated with RDA-like decimation steps. Different corrections to the neighboring defect cells $l$ and $r$ must be done for odd and even $M^{(i)}$. The iteration is repeated with the half-sized length until all ideal cells have been decimated. After final corrections and the treatment of all other periodic parts, the RGF can be applied.}
	\label{JCP1:fig:RGF+RDA:scheme}
\end{figure}

In this section we explain our improved approach, making the RGF, shown in the previous section, faster. We consider a central region with realistic defects like vacancies, substitutional atoms, and functionalization. In contrast to Anderson disorder, which randomizes the Hamiltonian matrix of the whole system, here, most of the sub-Hamiltonian matrices remain the ideal ones of the periodic system. Only a few sub-Hamiltonian matrices of the defective parts are different. Consequently, the total central region consists of (long) periodic parts which are connected through defective cells. The periodic parts can be treated effectively by decimating all the ideal cells using RDA steps. Especially in cases where the electrodes have the same structure as the defect-free parts of the central region, most of the computations have already been done during the electrode calculation. These decimations lead to corrected Hamiltonian matrices of the defects. Afterwards, the reduced Hamiltonian matrix consists only of corrected defects and can be treated by the original RGF. The computational complexity of the RDA-like part scales as $\order{\log N_\text{D}}$, where $N_\text{D}$ is the number of defects. The computational complexity of the RGF part scales as $\order{N_\text{D}}$.

\begin{figure}[tb]
	\centering
	\includegraphics{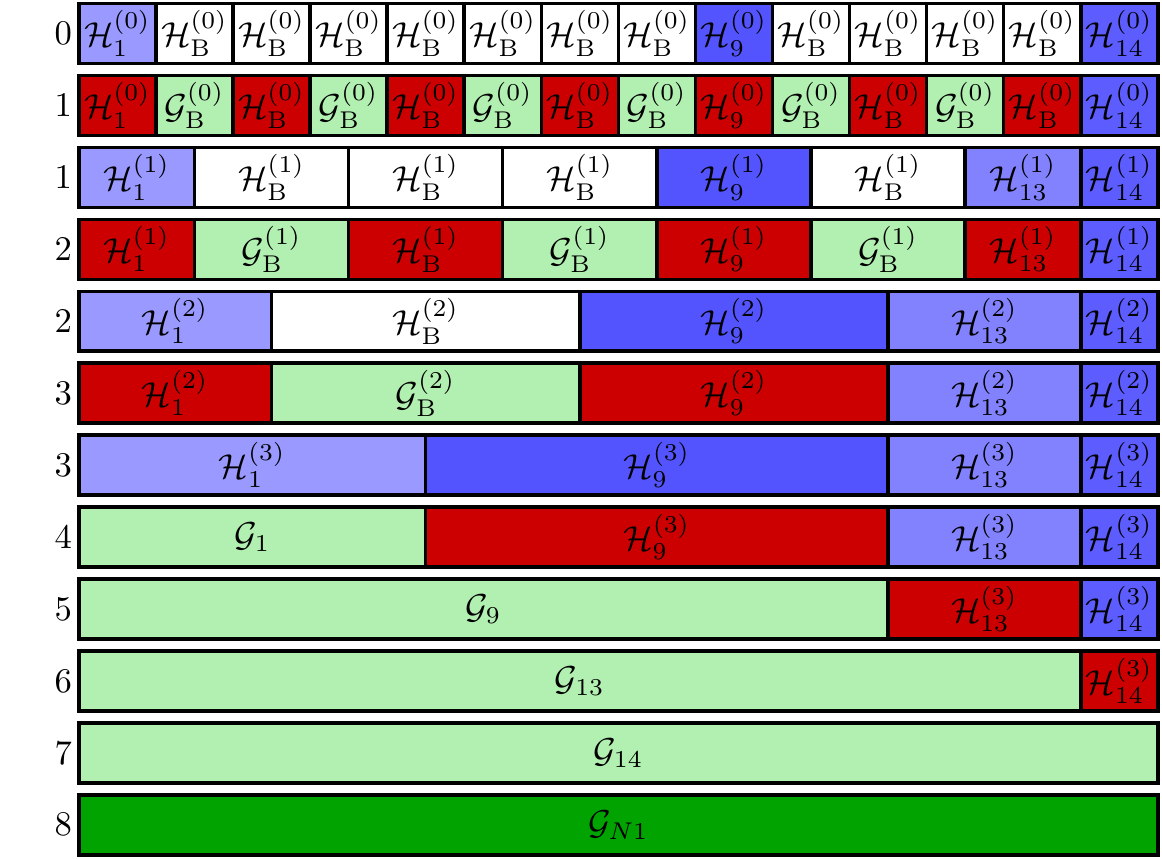}
	\caption[Sketch of the RGF+RDA]{(Color online.) Sketch of the RGF+RDA. From top to bottom: Hamiltonian matrices and Green's matrices when applying equations (\ref{JCP1:eqn:RGF-RDA:left}--\ref{JCP1:eqn:RGF-RDA:final}). The numbers at the left of each line denote the iteration step. Different blue blocks denote different Hamiltonian matrices (defect cells). White blocks denote same Hamiltonian matrices (ideal cells). Green blocks denote the calculation of the Green's matrix of the actual iteration step. The dark green block denotes the final step. Red blocks denote Hamiltonian matrices which are coupled to the highlighted Green's matrices. They get a self-energy correction. The first seven rows correspond to the RDA, the last six rows correspond to the RGF-FIS (compare figures \ref{JCP1:fig:RDA} and \ref{JCP1:fig:RGF}a).}
	\label{JCP1:fig:RGF+RDA}
\end{figure}

The RDA-like part is schematically summarized in figure \ref{JCP1:fig:RGF+RDA:scheme} and can be described as follows. Let us consider one of the periodic parts. Let $M$ be the number of lined up ideal cells. Let $l$ and $r=l+M+1$ be the indices of the two defect cells at the left and right end. At each step $i$, one decimates every second cell, starting with index $l+1$. As defect cell $l$ is connected to cell $l+1$, the corresponding Hamiltonian matrices and coupling matrices must be corrected in the following way:
\begin{articlesubequations}
	\label{JCP1:eqn:RGF-RDA:left}
	\begin{align}
		\hamilton_l^{(i+1)} &= \hamilton_l^{(i)} + \coupling_{l(l+1)}\green_\text{B}^{(i)}\coupling_{(l+1)l} \qquad,\label{JCP1:eqn:RGF-RDA:left:H}\\
		\coupling_{l(l+1)}^{(i+1)} &= \coupling_{l(l+1)}^{(i)}\green_\text{B}^{(i)}\alpha^{(i)} \qquad,\label{JCP1:eqn:RGF-RDA:left:Tupper}\\
		\coupling_{(l+1)l}^{(i+1)} &= \beta^{(i)}\green_\text{B}^{(i)}\coupling_{(l+1)l}^{(i)} \qquad.\label{JCP1:eqn:RGF-RDA:left:Tlower}
	\end{align}%
\end{articlesubequations}%
$\alpha$, $\beta$, and $\green_\text{B}$ are calculated via equation (\ref{JCP1:eqn:RDA}). If $M^{(i)}$ is odd, cell $r-1$ is decimated and the Hamiltonian matrices and coupling matrices of defect cell $r$ (which is connected to cell $r-1$) must be corrected in a similar way:
\begin{articlesubequations}
	\label{JCP1:eqn:RGF-RDA:right:odd}
	\begin{align}
		\hamilton_r^{(i+1)} &= \hamilton_r^{(i)} + \coupling_{r(r-1)}^{(i)}\green_\text{B}^{(i)}\coupling_{(r-1)r} \qquad,\label{JCP1:eqn:RGF-RDA:right:odd:H}\\
		\coupling_{(r-1)r}^{(i+1)} &= \alpha^{(i)}\green_\text{B}^{(i)}\coupling_{(r-1)r}^{(i)} \qquad,\label{JCP1:eqn:RGF-RDA:right:odd:Tupper}\\
		\coupling_{r(r-1)}^{(i+1)} &= \coupling_{r(r-1)}^{(i)}\green_\text{B}^{(i)}\beta^{(i)} \qquad.\label{JCP1:eqn:RGF-RDA:right:odd:Tlower}
	\end{align}%
\end{articlesubequations}%
If $M^{(i)}$ is even, cell $r-2$ is decimated. In this case, the last ideal cell gets only one correction term instead of two. It cannot be treated any more as an ideal cell within the RDA-like decimation and has to be assigned to the defect cells for further calculation. The corresponding corrections are
\begin{articlesubequations}
	\label{JCP1:eqn:RGF-RDA:right:even}
	\begin{align}
		r :=&~r-1 \qquad,\\
		\hamilton_r^{(i+1)} =&\; \hamilton_\text{R}^{(i+1)} \qquad,\label{JCP1:eqn:RGF-RDA:right:even:H}\\
		\coupling_{(r-1)r}^{(i+1)} =&\; \alpha^{(i+1)} \qquad,\label{JCP1:eqn:RGF-RDA:right:even:Tupper}\\
		\coupling_{r(r-1)}^{(i+1)} =&\; \beta^{(i+1)} \qquad.\label{JCP1:eqn:RGF-RDA:right:even:Tlower}
	\end{align}%
\end{articlesubequations}%
The last step, which decimates the last cell, yields $l+1=r$. Here, we also have to evaluate equations (\oldref{JCP1:eqn:RGF-RDA:left:H}) and (\oldref{JCP1:eqn:RGF-RDA:right:odd:H}), but we get the final effective coupling matrices by means of
\begin{articlesubequations}
	\label{JCP1:eqn:RGF-RDA:final}
	\begin{align}
		\coupling_{lr}^{(i+1)} &= \coupling_{l(l+1)}^{(i)}\green_\text{B}^{(i)}\coupling_{(r-1)r}^{(i)} \qquad,\label{JCP1:eqn:RGF-RDA:final:Tupper}\\
		\coupling_{rl}^{(i+1)} &= \coupling_{r(r-1)}^{(i)}\green_\text{B}^{(i)}\coupling_{(l+1)l}^{(i)} \qquad,\label{JCP1:eqn:RGF-RDA:final:Tlower}
	\end{align}%
\end{articlesubequations}%
instead of calculating equations (\oldref{JCP1:eqn:RGF-RDA:left:Tupper},\hyperref[JCP1:eqn:RGF-RDA:left:Tlower]{c};\oldref{JCP1:eqn:RGF-RDA:right:odd:Tupper},\hyperref[JCP1:eqn:RGF-RDA:right:odd:Tlower]{c}).

If $M$ is previously decomposed in the basis 2, that means $M=\sum_{i=0}^jM_i2^i$ with $M_j=1$, equations (\ref{JCP1:eqn:RGF-RDA:right:odd}) of level~$i$ (beginning with $i=0$ and ending with $i=j$) have to be executed if $M_i=1$ and equations (\ref{JCP1:eqn:RGF-RDA:right:even}) if $M_i=0$. Equations~(\ref{JCP1:eqn:RGF-RDA:left}) have always to be executed. For the final step, equations (\oldref{JCP1:eqn:RGF-RDA:left:H};\oldref{JCP1:eqn:RGF-RDA:right:odd:H};\ref{JCP1:eqn:RGF-RDA:final}) have to be executed.

Up to now, one periodic sequence is described. The other ones can be treated in the same way and independently. Thereafter, the remaining effective defect cells can be used as input for the RGF-FIS or the RGF-RDS. A sketch of the RGF+RDA is shown in figure \ref{JCP1:fig:RGF+RDA}.

The computational complexity of this algorithm cannot be easily obtained exactly and analytically, because the number of additional effective defect cells (and thus the number of matrix multiplications) is determined by the decomposition of $M$ into $M_i$'s and therefore it strongly depends on $M$ in a non-monotonous way for each sequence. The number of matrix inversions for a sequence of RDA steps, additional effective Hamiltonian matrix calculations, and RGF steps of ideal cells is $\mathcal{N}_\text{inv}(M)=2\left\lfloor\log_2M\right\rfloor$. The number of corresponding matrix multiplications is in the range $14\log_2\frac{M}{3}+35\leq\mathcal{N}_\text{mult}(M)\leq 19\log_2\frac{M+1}{3}+30$. Considering the whole partly-periodic system with randomly distributed defects, we will get an $\order{\log N_\text{D}}$ behavior for these RDA-like calculations (see section \ref{JCP1:testmatrix}). Afterwards, we are left with $N_\text{D}$ effective defect cells, which leads to an $\order{N_\text{D}}$ behavior for the RGF. The most important improvement is the logarithmic scaling behavior, stemming from the RDA-like treatment, in contrast to the linear scaling behavior of the RGF. A further advantage is that in cases where the electrodes are identical to the periodic sequences within the defective bulk part of the system, the RDA steps also occur in the electrode calculation, which further decreases the calculation time.

Finally, we want to make a remark about the generalization to two and three dimensions. For the common RGF this works by treating these systems in a quasi one-dimensional way: A two-dimensional system of $N_1\times N_2$ cells can be divided into $N_1$ stripes, which consist of $N_2$ cells. A three-dimensional system of $N_1\times N_2\times N_3$ cells can be divided into $N_1$ slices, which consist of $N_2\times N_3$ cells. The RGF can be applied to these $N_1$ stripes/slices, but their dimensions are then a factor $N_2$ resp. $N_2\times N_3$ larger. For the improvement for the case of randomly distributed realistic defects shown in this work, it would in principle be also possible to do so. But this means randomly distributed defective stripes/slices, which are separated by a large two-/three-dimensional area without defects, and randomly distributed defects within these stripes/slices. This does not describe the case of defects which are distributed randomly within the total two-/three-dimensional system. In other words: For a physical two-/three-dimensional system with randomly distributed defects a division into few small stripes/slices with defects cannot work. On the other hand, a cell-wise generalization does not work for the herein discussed real-space RGF, because the decimation of one cell causes coupling elements between all the cells, which are coupled to the decimated one. In one dimension, this transforms the block-tridiagonal matrix into a similar block-tridiagonal matrix. But in two/three dimensions, this does not transform the block-penta/heptadiagonal matrix into a similar block-penta/heptadiagonal matrix, because for each coupling element, which is removed, 7 resp. 26 coupling elements are added. The resulting algorithm would be worse than direct inversion.

\articlesection{Performance test}\label{JCP1:results}

In this section we evaluate the performance of the algorithm described in section \ref{JCP1:RGF-RDA}, which we have implemented in C++ using the LAPACK routines. We focus on calculation time (wall clock time) and the memory requirement (RAM).
We consider two systems: 1) A random test matrix to which we apply only the RGF+RDA. 2) An infinite carbon nanotube with randomly distributed divacancies to which we apply the whole transport formalism.

\articlesubsection{Random test matrix}\label{JCP1:testmatrix}

\begin{figure}[tb]
	\begin{minipage}{\OneColumnWidth}
		(a) $\dim\hamilton_i=64$
	\end{minipage}\hfill
	\begin{minipage}{\OneColumnWidth}
		(b) $\dim\hamilton_i=160$
	\end{minipage}\\[0.5em]
	\includegraphics{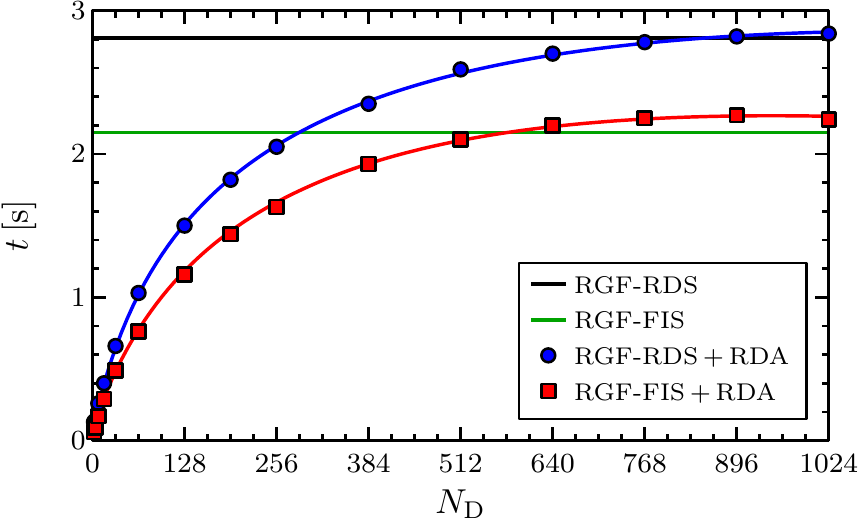}\hfill
	\includegraphics{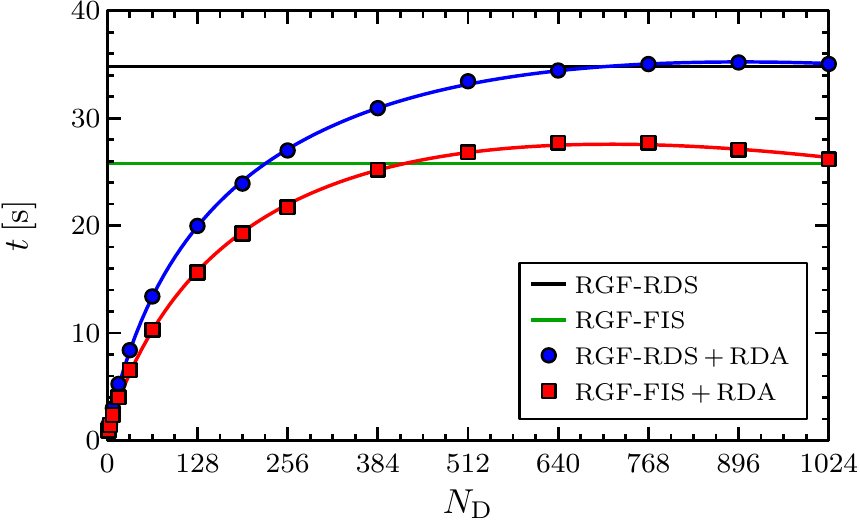}
	\caption[Calculation time for the RGF+RDA applied to random test matrices]{(Color online.) Calculation time for calculating $G_{N1}$ of a random test matrix which consists of 1024 blocks of dimension 64 (a) and 160 (b). $1024-N_\text{D}$ is the number of equal matrix blocks. $N_\text{D}$ is the number of deviant matrix blocks (defects). Colors denote the three different algorithms: the RGF-RDS (black), the RGF-FIS (green), and the RGF+RDA (blue and red). Dots and squares are data points. Lines are regressions according to the expected dependence.}
	\label{JCP1:fig:RGF:random_matrix}
\end{figure}

We compare the RGF+RDA steps with the common RGF-FIS and RGF-RDS regarding calculation time and its dependence on the defect fraction. For this purpose, we construct a test matrix, which is block-tridiagonal. The matrix blocks are either ideal ones or defective ones. The ideal matrix blocks are all equal, chosen as a random complex matrix. The defect matrix blocks are all different random complex matrices.

Figure \ref{JCP1:fig:RGF:random_matrix} shows the calculation time as a function of the number of defect matrix blocks $N_\text{D}$. The total test matrix consists of $N=1024$ blocks of dimension $\dim\hamilton_i=64$ (figure \ref{JCP1:fig:RGF:random_matrix}a) and $\dim\hamilton_i=160$ (figure \ref{JCP1:fig:RGF:random_matrix}b). This is comparable to the respective dimensions of a (4,4)- and a (10,10)-CNT, which we discuss later. The calculation time of the RGF-FIS and the RGF-RDS is independent of the number of defect matrix blocks. The difference between the FIS and the RDS is the amount of matrix multiplications. The RDS needs $3N-6$ multiplications more than the FIS.

For low defect fractions, the improved RGF+RDA scales as $\log(N_\text{D})$, according to the RDA part, and as $N_\text{D}$, according to the RGF part. At $N_\text{D}=0$ it results in a pure RDA. Likewise, at $N_\text{D}=N$ it results in a pure RGF. Note that for 1024 defect-free cells, a pure RDA needs nearly no time in comparison to a pure RGF. For very high defect fractions ($>0.5$), the combined approach requires a bit more calculation time than the RGF alone. This is due to the fact that the RDA-like treatment needs more matrix multiplications for the Hamiltonian matrix corrections than one RGF step. This plays a role, especially, for few and short periodic parts. The calculation time increase, caused by the additional matrix multiplications, is higher than the calculation time reduction, caused by the reduced number of matrix inversions. For fixed $N$, the overall calculation time of the RGF+RDA can be described by $t=a+bN_\text{D}+c\log(d+N_\text{D})$ with specific constants $a$, $b$, $c$, $d$, in contrast to a constant time within a pure RGF treatment.

\articlesubsection{Transport through carbon nanotubes}\label{JCP1:cnt}

\begin{figure}[tb]
	\centering
	\begin{minipage}{0.75\textwidth}
		\newlength{\mylength}
		\setlength{\mylength}{0.1\textwidth}
		\begin{minipage}{0.5\mylength}
			~
		\end{minipage}\hfill
		\begin{minipage}{\mylength}
			\centering
			UC
		\end{minipage}\hfill
		\begin{minipage}{2.4\mylength}
			\centering
			DV$_\text{perp}$
		\end{minipage}\hfill
		\begin{minipage}{2.4\mylength}
			\centering
			DV$_\text{diag1}$
		\end{minipage}\hfill
		\begin{minipage}{3\mylength}
			\centering
			DV$_\text{diag2}$
		\end{minipage}\\[0.3em]
		\begin{minipage}{0.5\mylength}
			\centering
			\begin{sideways}
				(4,4)-CNT
			\end{sideways}
		\end{minipage}\hfill
		\begin{minipage}{\mylength}
			\includegraphics[width=\textwidth]{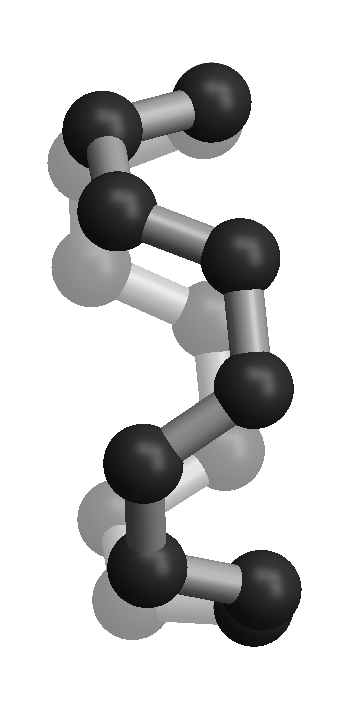}
		\end{minipage}\hfill
		\begin{minipage}{2.4\mylength}
			\includegraphics[width=\textwidth]{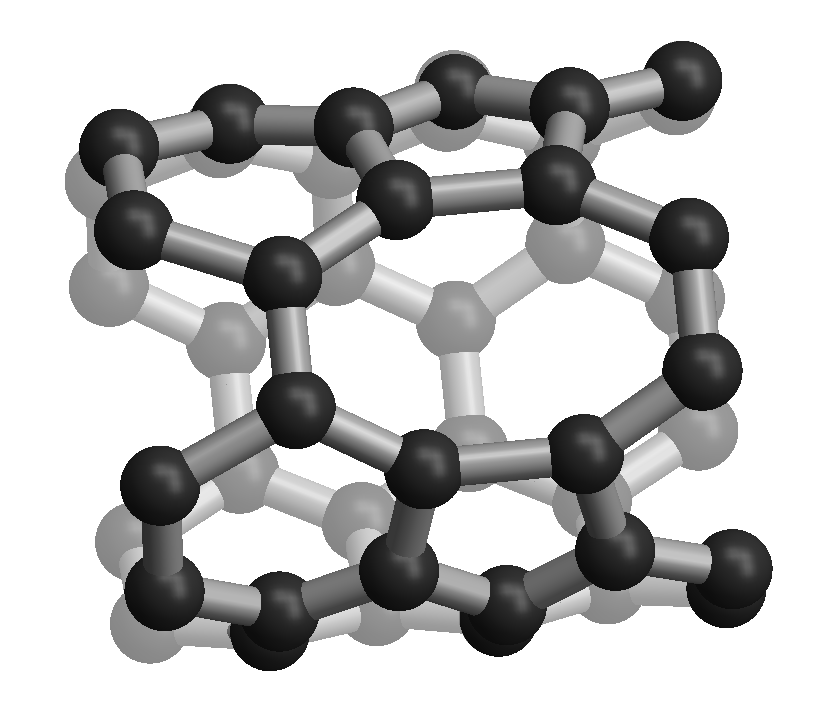}
		\end{minipage}\hfill
		\begin{minipage}{2.4\mylength}
			\includegraphics[width=\textwidth]{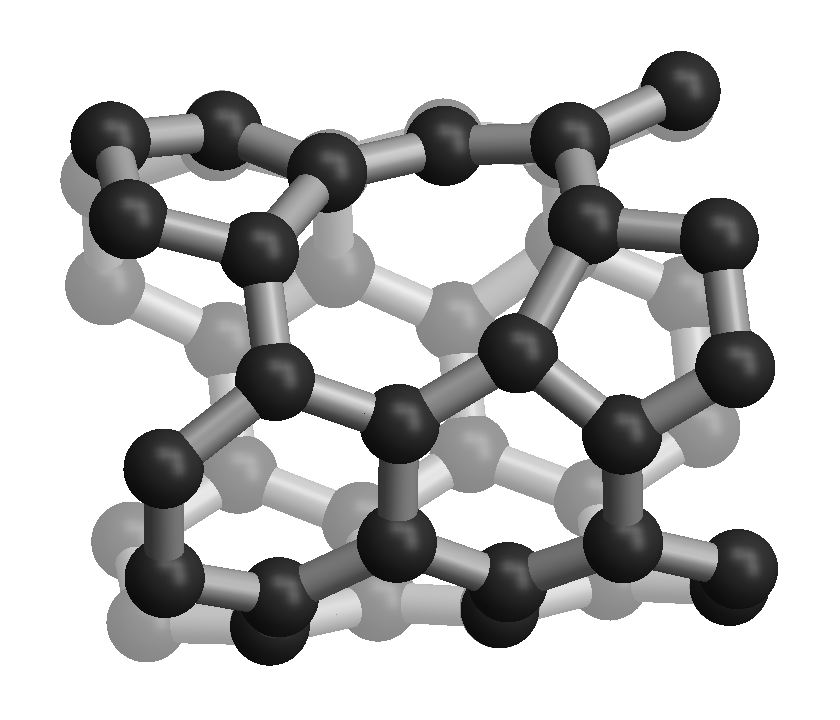}
		\end{minipage}\hfill
		\begin{minipage}{3\mylength}
			\includegraphics[width=\textwidth]{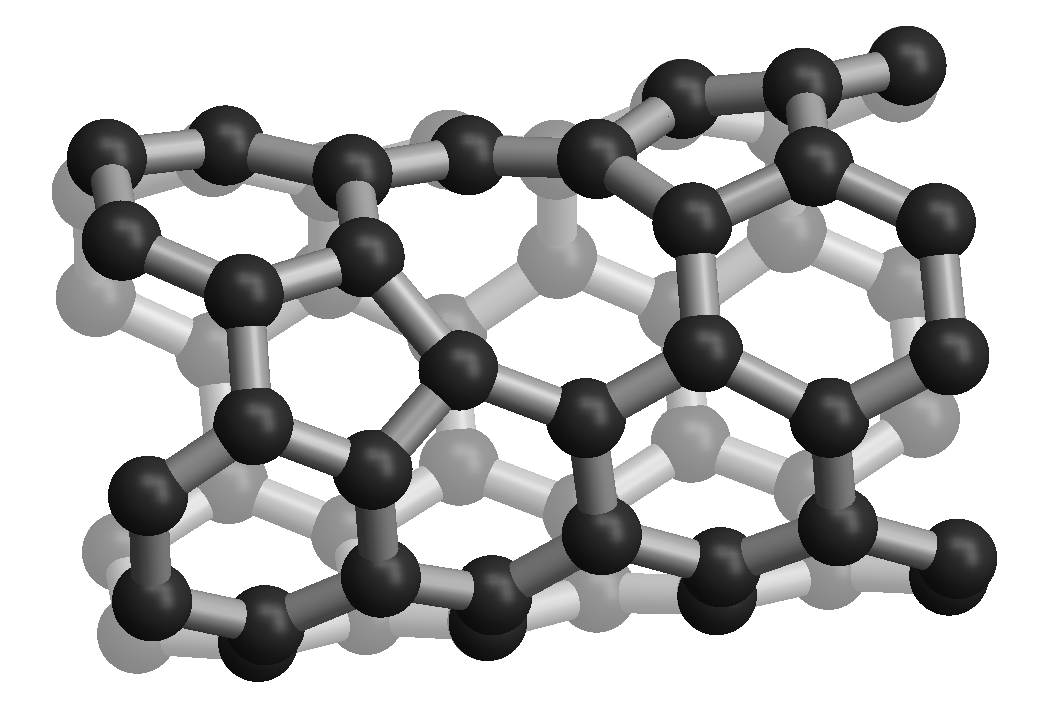}
		\end{minipage}\\[0.3em]
		\begin{minipage}{0.5\mylength}
			\centering
			\begin{sideways}
				(10,10)-CNT
			\end{sideways}
		\end{minipage}\hfill
		\begin{minipage}{\mylength}
			\includegraphics[width=\textwidth]{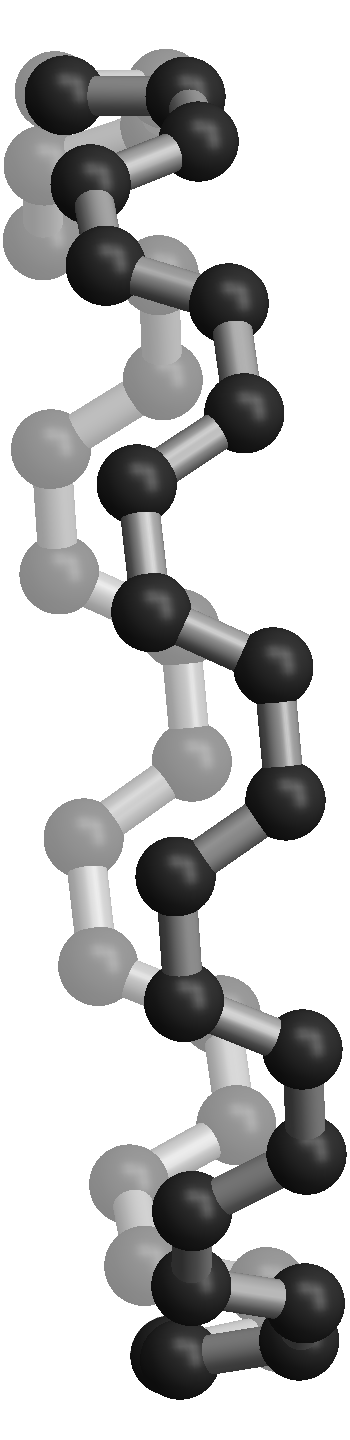}
		\end{minipage}\hfill
		\begin{minipage}{2.4\mylength}
			\includegraphics[width=\textwidth]{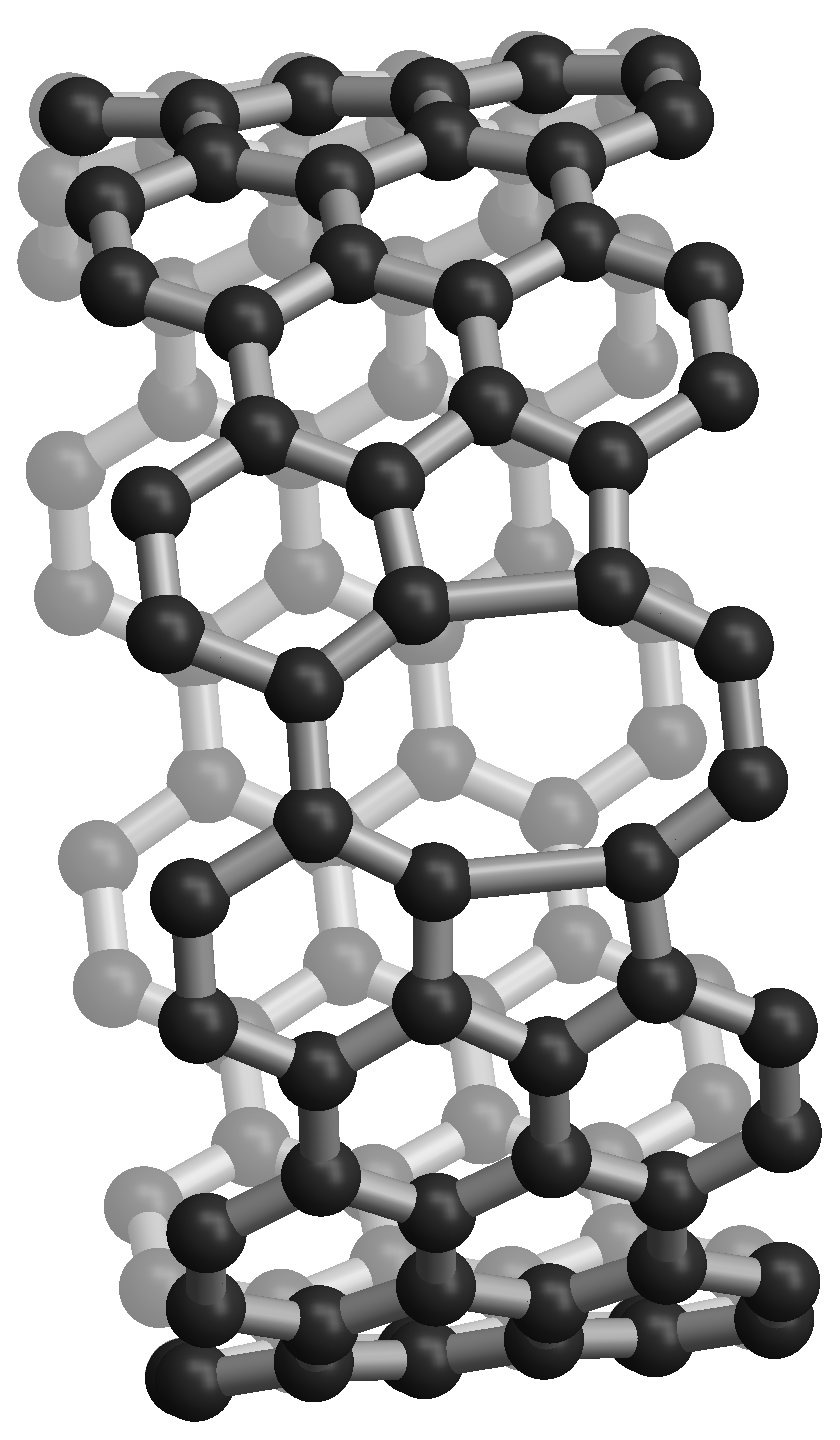}
		\end{minipage}\hfill
		\begin{minipage}{2.4\mylength}
			\includegraphics[width=\textwidth]{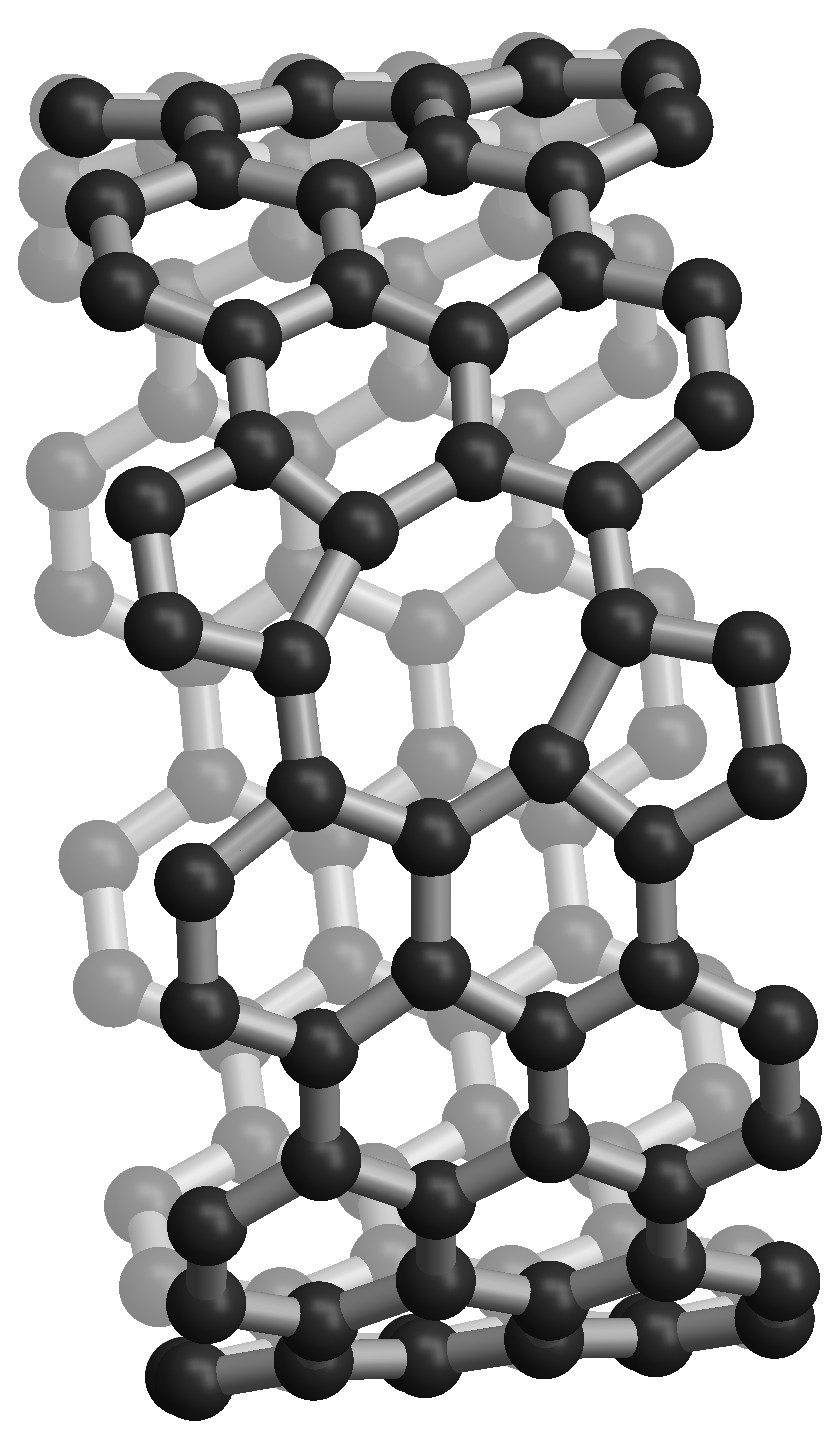}
		\end{minipage}\hfill
		\begin{minipage}{3\mylength}
			\includegraphics[width=\textwidth]{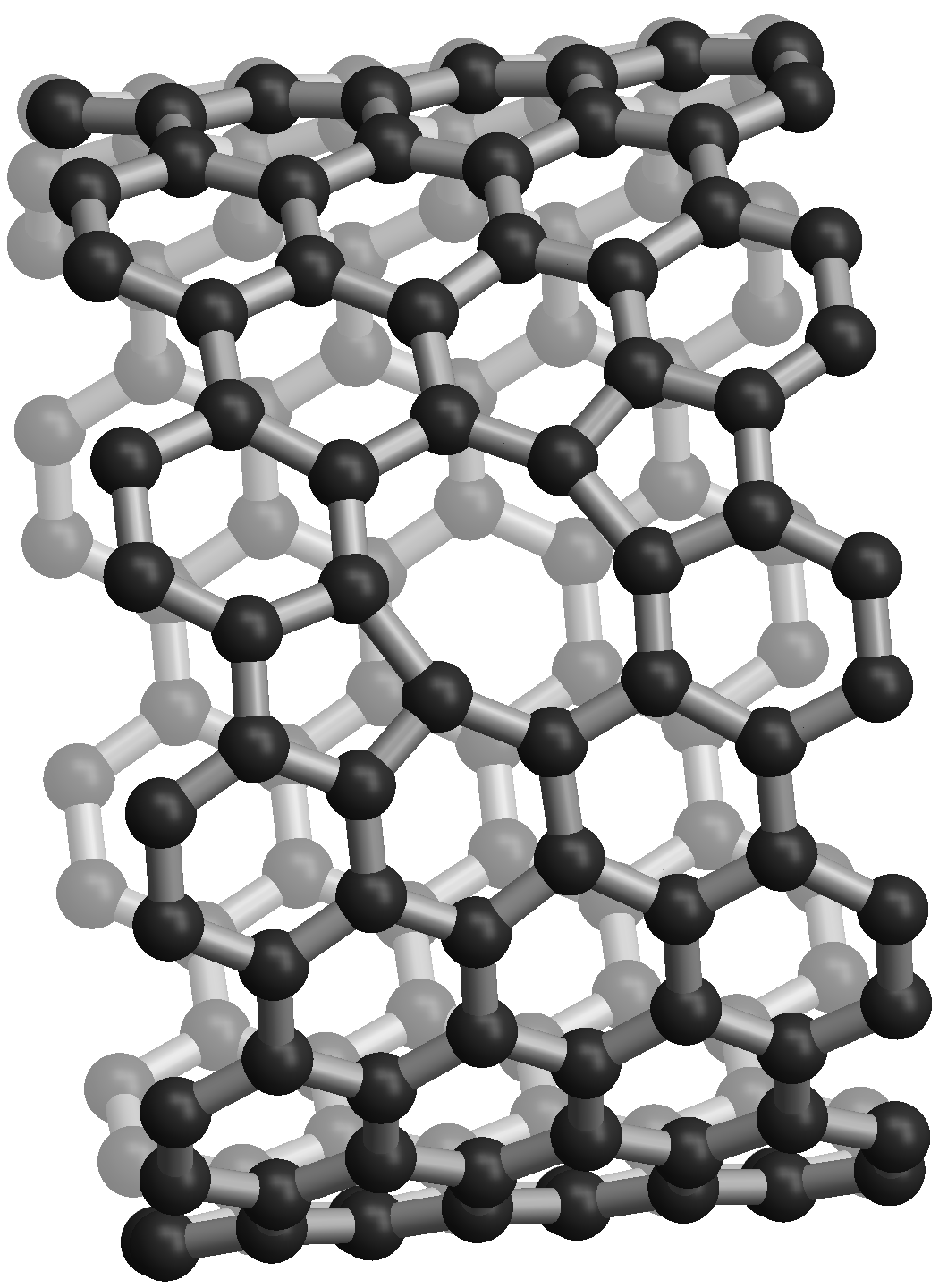}
		\end{minipage}
	\end{minipage}
	\caption[Geometric structures of the CNT cells]{Atomic structure of the used cells of the (4,4)-CNT (upper row) and the (10,10)-CNT (lower row), from left to right: unit cell of the ideal CNT (UC), divacancy with perpendicular orientation (DV$_\text{perp}$), and the two types of the divacancy with diagonal orientation (DV$_\text{diag}$). Each subsystem of the RGF (figure \ref{JCP1:fig:transport:scheme}b) is chosen as one of these cells.}
	\label{JCP1:fig:cnt:geometry}
\end{figure}

\begin{figure}[!tb]
	(a) calculation time\\[0.5em]
	\includegraphics{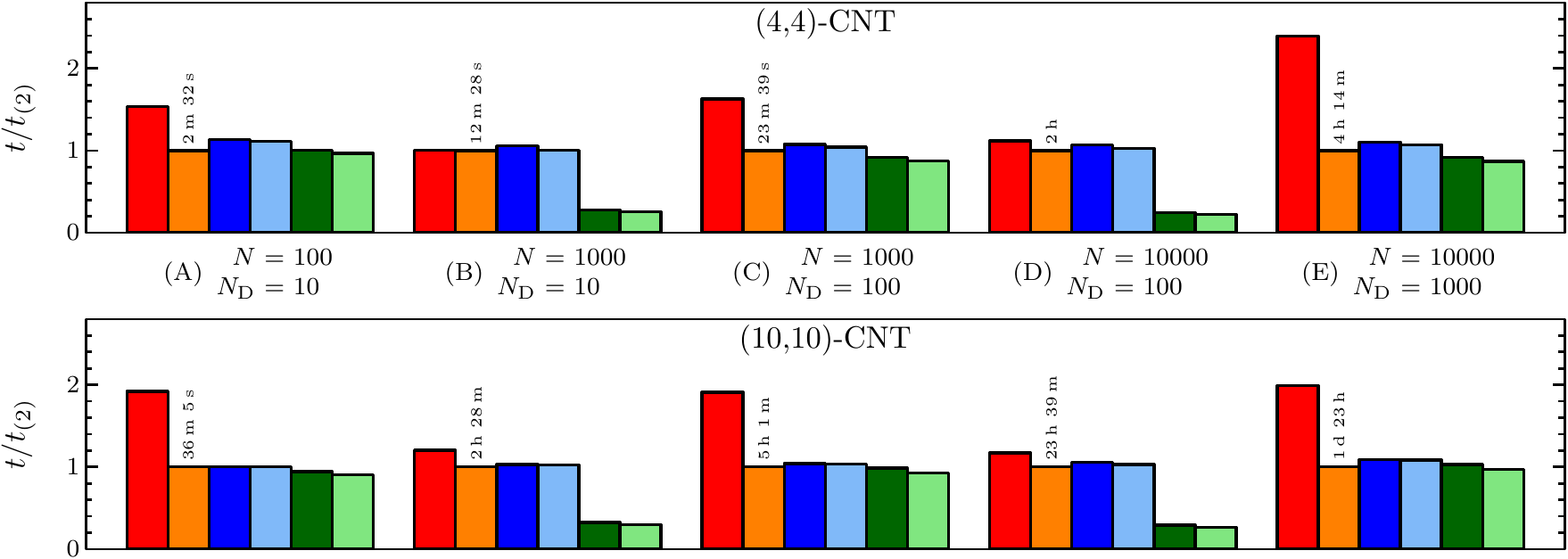}\\[0.5em]
	(b) memory usage\\[0.5em]
	\includegraphics{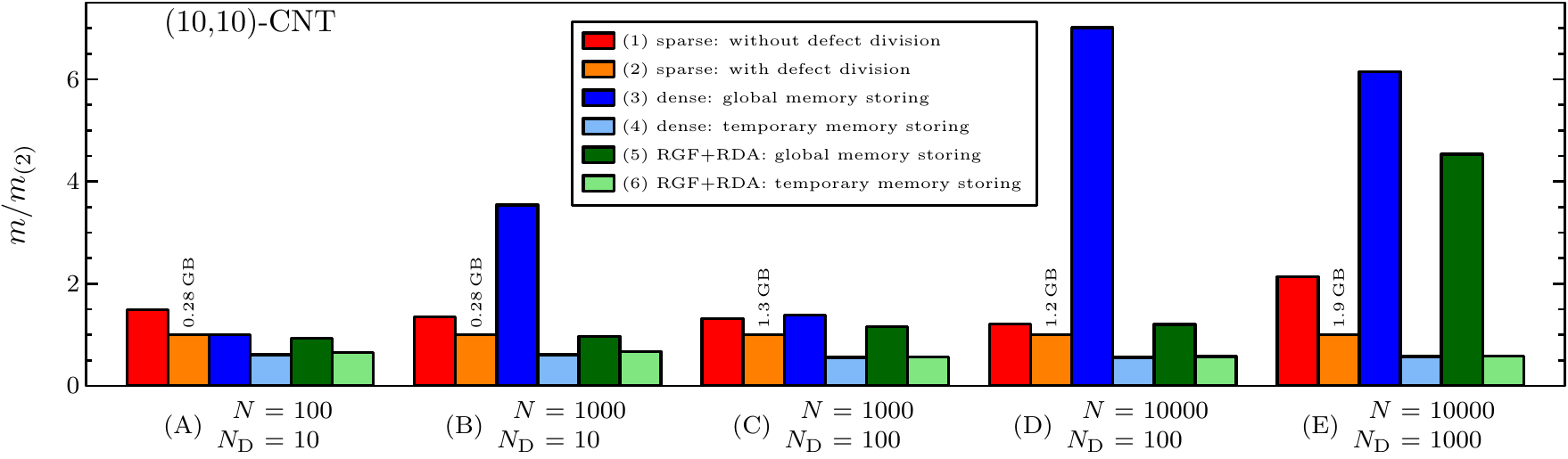}
	\caption[Calculation time and memory usage for the RGF+RDA applied to CNTs]{(Color online.) (a) Calculation time for defective (4,4)-CNTs (upper diagram) and (10,10)-CNTs (lower diagram). (b) Memory usage for defective (10,10)-CNTs. Each diagram shows five different examples (A)\ldots(E) with different number of CNT cells~$N$ and defect cells~$N_\text{D}$. Different colors denote different algorithms and matrix treatment (see text for detailed explanation). The values are normalized to case (2), the sparse format with defect division. The absolute values are given above the corresponding bars.}
	\label{JCP1:fig:RGF:cnt}
\end{figure}

To study a more realistic system, we choose armchair-(4,4)- and (10,10)-CNTs. We focus on divacancies, which are the most common defects in CNTs~\cite{PhysRevB.63.245405, NanoLett.9.2285, JPhysDApplPhys.43.305402}. Figure \ref{JCP1:fig:cnt:geometry} shows the atomic structure of their unit cells (UCs) and different orientations of divacancies (DV) within the respective tubes. The unit cell consists of 16 atoms for the (4,4)-CNT and 40 atoms for the (10,10)-CNT. The divacancy defect has the same structure, but with two adjacent atoms removed. There are three different types, labeled DV$_\text{perp}$ (aligned perpendicular to the tube axis) and DV$_\text{diag1}$/DV$_\text{diag2}$ (aligned diagonal). Because of rotational symmetry, each type comprises 8 positions for the (4,4)-CNT and 20 positions for the (10,10)-CNT (which are not shown in the figure). All defect cells are much longer than one unit cell to capture the locally distorted atomic structure, caused by the defect. It results in 46 (DV$_\text{perp}$, DV$_\text{diag1}$) or 62 (DV$_\text{diag2}$) atoms per defect cell for the (4,4)-CNT and 118 (DV$_\text{perp}$, DV$_\text{diag1}$) or 158 (DV$_\text{diag2}$) atoms for the (10,10)-CNT. The geometric structures have been generated by a geometry optimization with density functional theory~\cite{BrazJPhys.36.1318}, as implemented in Atomistix ToolKit~\cite{ATK.12.8.2, PhysRevB.65.165401}. We use the local density approximation of Perdew and Zunger~\cite{PhysRevB.23.5048}, norm-conserving Troullier-Martins pseudopotentials \cite{PhysRevB.43.1993}, and a double zeta plus double polarization (DZDP) basis set of the SIESTA type \cite{JPhysCondMat.14.2745}.

The electronic structure (i.e. the Hamiltonian matrices and the coupling matrices) is obtained by a density-functional-based tight-binding (DFTB) model~\cite{PhysRevB.51.12947, IntJQuantumChem.58.185}. We use the existing parameter set 3ob~\cite{JChemTheoryComput.9.338, PhysRevB.58.7260}, which contains onsite energies, distance-dependent hopping energies, and overlap elements for carbon within a four-orbital sp$^3$-basis. To reduce the size of the cells, an interaction cutoff distance of twice the carbon-carbon distance was chosen, which is a good compromise between calculation efficiency and accuracy~\cite{NJPhys.16.123026}. This leads to a distance-dependent third-nearest-neighbor description, allowing us to use two carbon rings for the unit cell.

Figure \ref{JCP1:fig:RGF:cnt} shows the calculation time and the memory consumption of the whole transport formalism. Besides the calculation of $\green_{N1}$, this includes the memory allocation time, some additional calculations to reduce memory requirements, the calculation of the electrodes with the RDA to get $\varSigma_\text{L/R}$ and $\varGamma_\text{L/R}$, the calculation of the energy-dependent transmission function using (\ref{JCP1:eqn:T(E):RGF}), and the calculation of $\transmission(E)$ after every defect to get the length dependence. We consider five examples with different numbers of cells~$N$ and defect cells~$N_\text{D}$:
\begin{enumerate}
	\item[(A)] \num{100} cells with \num{10} defects ($p_\text{D}=0.1$),
	\item[(B)] \num{1000} cells with \num{10} defects ($p_\text{D}=0.01$),
	\item[(C)] \num{1000} cells with \num{100} defects ($p_\text{D}=0.1$),
	\item[(D)] \num{10000} cells with \num{100} defects ($p_\text{D}=0.01$), and
	\item[(E)] \num{10000} cells with \num{1000} defects ($p_\text{D}=0.1$).
\end{enumerate}
The defect fraction is either $p_\text{D}=0.1$ (in A, C, E) or $p_\text{D}=0.01$ (in B, D). Furthermore, we distinguish six different cases according to the algorithm and the matrix treatment:
\begin{enumerate}
	\item[(1)] Defect cells are not divided. The matrices are stored globally in a sparse format. The RGF-FIS (section \ref{JCP1:FIS}) is used.
	\item[(2)] Defect cells are divided. The matrices are stored globally in a sparse format. The RGF-FIS is used.
	\item[(3)] Defect cells are divided. The matrices are stored globally in a dense format. The RGF-FIS is used.
	\item[(4)] Defect cells are divided. The matrices are stored temporarily. The RGF-FIS is used.
	\item[(5)] Defect cells are divided. The matrices are stored globally in a dense format. The RGF-FIS+RDA approach (section \ref{JCP1:RGF-RDA}) is used.
	\item[(6)] Defect cells are divided. The matrices are stored temporarily. The RGF-FIS+RDA approach is used.
\end{enumerate}
For better comparison the calculation time and the memory consumption shown in figure \ref{JCP1:fig:RGF:cnt} are normalized to case (2). The corresponding absolute values are given above the respective bars.

First, we want to mention a trivial calculation time reduction, comparing (1) and (2). The cell-wise RGF treatment suggests taking one defect cell in one iteration step (1), but all defects are longer than the ideal UCs due to the extension of the distorted structure. Consequently, the corresponding defect Hamiltonian matrix will also be block-tridiagonal and thus, it can be further divided. Concerning our examples, DV$_\text{perp}$ and DV$_\text{diag1}$ are subdivided into two cells and DV$_\text{diag2}$ into three cells in algorithm (2)\footnote{Although DV$_\text{perp}$ and DV$_\text{diag1}$ have the same length as three UCs, they are subdivided only into two cells, because of the following fact: The interaction cutoff distance was set to twice the carbon-carbon distance, which is equal to the third-nearest-neighbor distance of carbon atoms. At the same time, the UC length is equal to the second-nearest-neighbor distance of carbon atoms, which is allowed, because a coupling between second-nearest-neighbor cells $\coupling_{i(i+2)}$ needs an interaction between fourth-nearest-neighbor carbon atoms. Due to the distorted carbon rings in the divacancy structures (see figure \ref{JCP1:fig:cnt:geometry}), a subdivision of DV$_\text{perp}$ and DV$_\text{diag1}$ into three cells could cause a second-nearest-neighbor coupling $\coupling_{i(i+2)}$, which has to be prevented by dividing DV$_\text{perp}$ and DV$_\text{diag1}$ into two cells instead of three. This also applies to the subdivision of DV$_\text{diag2}$ into three cells instead of four.}. Such a defect division always makes sense as it reduces the calculation time of the corresponding RGF steps in comparison to non-divided defect cells, as can be seen by comparing (1) and (2) in figure \ref{JCP1:fig:RGF:cnt}a.

We implemented two different ways of storing matrices: matrix blocks for different cell types and coupling types are stored once and one of these types is assigned to every cell of the CNT sequence. We call it sparse format. This allows us to do some calculations at the beginning and use the results later, instead of repeating identical computations during the RGF steps. Thus, this should result in a more effective algorithm in comparison to storing the matrix blocks of all cells of the CNT sequence, what we call dense format. In figure \ref{JCP1:fig:RGF:cnt}a we see the reduced calculation time of (2) compared to (3). The reduction lies in the range of 5\% to 10\% for long systems. The corresponding memory consumption is visualized in figure \ref{JCP1:fig:RGF:cnt}b. The global matrix storage (3) leads to an extensive memory usage, exceeding the acceptable limit of computer resources. But since each iteration step only needs parts of the overall Hamiltonian matrix, they can be loaded temporarily and deleted afterwards. This variant is denoted by (4). The memory usage is reduced drastically and is even better than the one of the sparse format (1, 2). The difference of the matrix storage also affects the calculation time. Since the reservation of memory is lower for the temporary matrix storing (4), also the calculation time is slightly reduced. This reduces the calculation time advantages of using the sparse format (2) to one half.

The most relevant part of figure \ref{JCP1:fig:RGF:cnt}a is the comparison between (2) and the new RGF+RDA (5) and (6). Note that the improved RGF+RDA cannot be used in combination with the sparse matrix storage because the effective defect matrix blocks are affected differently by the periodic parts in between. For the global matrix storage (5) and the high defect probability $p_\text{D}=0.1$ (examples A, C, E), the calculations take nearly the same time. However, for the low defect probability $p_\text{D}=0.01$ (examples B, D), there is a large reduction of calculation time in the range of 67\% to 75\%. The calculation time reduction for the (10,10)-CNT is always smaller than for the (4,4)-CNT because of the much larger Hamiltonian matrix blocks. The temporary matrix storage (6) affects the required memory and the calculation time in a similar way as for the dense format. The comparison (6) vs. (5) in figure \ref{JCP1:fig:RGF:cnt}a shows a small decrease of the calculation time. This effect is at most 5\%.

In summary, the usage of the RGF+RDA approach with temporary matrix storage (6) is always advantageous compared to the pure RGF (2). It needs less memory (if matrices are only stored temporarily) and it is faster than the common RGF approach. We get calculation time reductions of at least about 5\% for large defect probabilities of $p_\text{D}=0.1$, and up to 80\% for $p_\text{D}=0.01$. Finally, we want to mention that the results of our calculations, which are not shown in this paper, have been published in \cite{NJPhys.16.123026}. Therein, we presented the conductance of CNTs with monovacancy defects and divacancy defects, and discussed its dependence on various parameters comprehensively.

\articlesection{Summary and conclusions}\label{JCP1:conclusions}

We developed an improved quantum transport algorithm for quasi one-dimensional devices with few realistic defects. The block-tridiagonal Hamiltonian matrix for such devices has long periodic parts, which are interrupted by few defect blocks. The improved RGF+RDA combines the RGF-FIS with the RDA. The periodic parts are treated by the RDA taking advantage of the previously performed electrode calculations. For a fixed number of cells, the scaling behavior of the computational complexity of this new approach has two parts: the RDA part scales logarithmically with the number of defects and the RGF part scales linearly with the number of defects. In contrast to this, the pure RGF would scale linearly with the total number of cells. Overall, this yields a reduction of the computational complexity of the RGF+RDA approach.

The logarithmic scaling behavior is especially advantageous for small defect probabilities, as shown for a test system of random matrices. Applying the algorithm to a more realistic system of practical interest, carbon nanotubes with divacancy defects, the overall calculation time is reduced by up to 80\% for $p_\text{D}=0.01$. At that, temporary data loading prevents excessive memory demands (in contrast to global data loading at the beginning), while computation times are nearly unaffected.

This work contributes to the continuing development of numerical implementations in quantum transport theory. Exploring possibilities to unify or combine different approaches, as demonstrated in the present paper, continues to be an important topic for future studies. The RGF+RDA is not limited to CNTs. It can be used for all quasi one-dimensional materials with realistic defects and low defect densities like graphene nanoribbons or nanowires. It is also applicable in a straightforward way to branched systems~\cite{JComputPhys.261.256}. Besides that, further improvements towards more arbitrary geometries, e.g. non-periodic edges of graphene nanoribbons or defects in 2D materials would be interesting and promising.

\articlesection*{Acknowledgement}

This work is funded by the European Union (ERDF) and the Free State of Saxony via the ESF project 100231947 (Young Investigators Group Computer Simulations for Materials Design - CoSiMa).
\begin{center}
	\centering\includegraphics[width=0.4\textwidth]{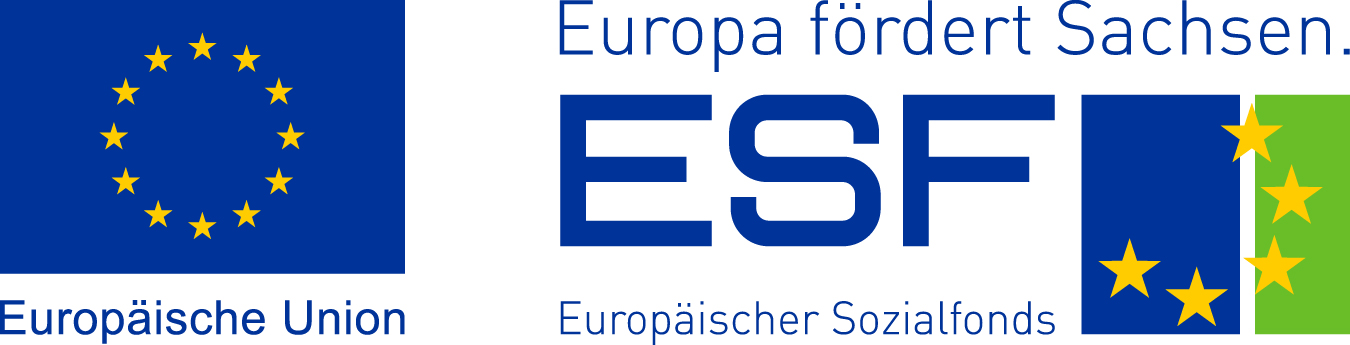}
\end{center}

\end{document}